\documentclass[aps, prd, twocolumn, superscriptaddress, nofootinbib]{revtex4}
\usepackage{graphicx}
\usepackage{dcolumn}
\usepackage{amssymb}

\begin{document}

\newcommand{\Eq}[1]{\mbox{Eq. (\ref{eqn:#1})}}
\newcommand{\Fig}[1]{\mbox{Fig. \ref{fig:#1}}}
\newcommand{\Sec}[1]{\mbox{Sec. \ref{sec:#1}}}

\newcommand{\PHI}{\phi}
\newcommand{\PhiN}{\Phi^{\mathrm{N}}}
\newcommand{\vect}[1]{\mathbf{#1}}
\newcommand{\Del}{\nabla}
\newcommand{\unit}[1]{\;\mathrm{#1}}
\newcommand{\x}{\vect{x}}
\newcommand{\ScS}{\scriptstyle}
\newcommand{\ScScS}{\scriptscriptstyle}
\newcommand{\xplus}[1]{\vect{x}\!\ScScS{+}\!\ScS\vect{#1}}
\newcommand{\xminus}[1]{\vect{x}\!\ScScS{-}\!\ScS\vect{#1}}
\newcommand{\diff}{\mathrm{d}}

\newcommand{\be}{\begin{equation}}
\newcommand{\ee}{\end{equation}}
\newcommand{\bea}{\begin{eqnarray}}
\newcommand{\eea}{\end{eqnarray}}
\newcommand{\vu}{{\mathbf u}}
\newcommand{\ve}{{\mathbf e}}


\title{MONDian three-body predictions for LISA Pathfinder}

\newcommand{\addressImperial}{Theoretical Physics, Blackett Laboratory, Imperial College, London, SW7 2BZ, United Kingdom}

\author{Neil Bevis} 
\email{n.bevis@imperial.ac.uk}
\affiliation{\addressImperial}

\author{Jo\~{a}o Magueijo} 
\email{magueijo@ic.ac.uk}
\affiliation{\addressImperial}

\author{Christian Trenkel} 
\email{Christian.Trenkel@astrium.eads.net}
\affiliation{Astrium Ltd, Gunnels Wood Road, Stevenage SG1 2AS, United Kingdom}

\author{Steve Kemble} 
\affiliation{Astrium Ltd, Gunnels Wood Road, Stevenage SG1 2AS, United Kingdom}

\date{\today}

\begin{abstract}
In previous work it was shown that MOND theories predict 
anomalously strong tidal
stresses near the saddle points of the Newtonian gravitational potential. An
analytical examination of the saddle between two bodies revealed
a linear and a non-linear solution, valid for the outer and inner regions. 
Here we present a numerical algorithm for solving the MOND equations. We
check the code against the two-body analytical solutions and explore the region
transitioning between them. We then develop a 
a realistic model for the MONDian effects on the saddles of the 
Sun-Earth-Moon system (including further sources is straightforward). 
For the Sun-Earth saddle we find that the two-body results 
are almost unchanged, with corrections increasing from 
full to new Moon. In contrast, the Moon saddle is an intrinsically 
three-body problem, but we numerically find a recipe for adapting
the two-body solution to this case, by means of a suitable re-scaling and axis
re-orientation.
We explore possible experimental scenarios for LISA Pathfinder,
and the prospect of a visit to the saddle(s) at the end of the mission.
Given the chaotic nature of the orbits, awareness of the full
range of the possibilities is crucial for a realistic prediction. 
We conclude that 
even with very conservative assumptions
on the impact parameter, the accelerometers are abundantly sensitive
to vindicate or rule out the theory.
\end{abstract}


\maketitle


\section{Introduction}
\label{sec:intro}

Modified Newtonian dynamics (MOND~\cite{Milgrom:1983ca}) is a scheme
that was first proposed for explaining observed dynamical properties of galaxies
without invoking dark matter.  The scheme was later incorporated into a 
proper theory with a Lagrangian formulation~\cite{aqual}, but valid 
only in the non-relativistic regime. Still later
a fully covariant gravitational theory was found containing MOND
phenomenology as a non-relativistic limit. This theory was named
T$e$V$e$S~\cite{teves} and alternatives 
have been proposed (e.g.~\cite{BSTV,aether,aether1}). 
The observational features to be studied in this paper 
depend only on their (shared) non-relativistic limit. 
Constraints arising from lensing~\cite{yusaf,yusaf1,yusaf2},
or cosmology~\cite{kostasrev}
have no bearing here and indicate issues arising from 
the relativistic extension of these theories.

The opposition between MOND and dark matter 
leaves considerable doubts as to how to interpret new astrophysical and
cosmological data. A fair comparison requires re-evaluating,
within each approach, 
the whole set of assumptions underlying the new observations
(see for example the controversy surrounding the 
bullet cluster~\cite{bullet,bullet1,bullet2,bullet3}).
For this reason the debate would benefit from a direct probe,
in the form of a laboratory 
or Solar System experiment.
Such a perspective motivates widespread dark matter searches. 
The analogous ``backyard'' tests of MOND include 
searching for anomalies in planetary and spacecraft 
trajectories~\cite{Sereno,pioneer0,pioneer}, stronger tidal 
stresses in the vicinity of saddle points of the 
Newtonian potential~\cite{Bekenstein:2006fi} or Solar System manifestations of 
the MOND external field effect~\cite{Milgromss}. 

In this paper we focus on the MONDian effects on the saddle points
of the gravitational field, 
with particular emphasis on the region where the  
Earth and Sun pulls cancel (which, we stress, is not at the Lagrange 
point L1). We provide realistic predictions for 
the future LISA Pathfinder spacecraft~\cite{LISA}, which could 
plausibly be redirected to the Earth-Sun saddle, once its primary goals 
at L1 have been completed. We also consider the possible benefits of 
re-direction to the saddle point near the Moon (there 
are not two separate Moon-Earth and Moon-Sun saddle points) and discuss 
its related practical issues.

As already mentioned, the predictions we calculate stem from the 
non-relativistic limit of T$e$V$e$S~\cite{teves}. For this theory gravity 
is described by the total potential $\Phi=\PhiN+\PHI$, where $\PhiN$ 
is the Newtonian potential and $\phi$ is an additional MOND component. 
The latter is physically relevant 
only when $|\Del \PhiN| \alt a_{0}$, where $a_{0}$ 
is the Milgrom acceleration~\cite{Milgrom:1983ca}, with
$a_0\approx 10^{-10}\unit{ms^{-1}}$ in order to fit galaxy observations 
without dark matter. The extra field is governed by the non-linear 
Poisson equation:
\begin{equation}
\label{eqn:TeVeS}
\vect{\Del}\cdot\left[ \mu \left(\kappa \left| \vect{\Del}\PHI \right| / a_{0} \right) \vect{\Del}\PHI \right] = \kappa G\rho
\end{equation}
where $\rho$ is the matter density, $\kappa$ is a constant parameter and $G$ is the 
underlying gravitational constant. On the left-hand-side, $\mu(y)$ is 
an unknown function that must tend to $1$ for $y\gg1$ 
but behaves like $y$ for $y\ll1$. 
We stress that this function is not the ratio of the Newtonian and actual 
accelerations, which we will denote as $\tilde{\mu}(x)$ with $x=a/a_0$,
a function more often employed by the astronomy community. Although for 
spherically symmetric cases the conversion between $\tilde{\mu}(x)$ 
and $\mu(y)$ is straightforward, in more general situations complications
may arise (see~\cite{kostasrev}). We will choose a particular form for $\mu(y)$ 
following Ref.~\cite{Bekenstein:2006fi} (see Eq.~\ref{eqn:mu} in Section~\ref{sec:twoBody} of the present paper), and likewise take $\kappa=0.03$ throughout this article. We'll comment on other choices of $\mu$ and their implications
for our results, in the concluding Section of this paper. 

In regions where $|\Del \PhiN| \gg a_{0}$, we have 
$\mu\approx1$ and therefore $\phi$ yields accelerations that are 
$\kappa/4\pi$ times the Newtonian contribution. Hence by measuring the total 
gravitational force we would measure the Newtonian gravitational constant 
to be $(1+\kappa/4\pi)G$ and \emph{not} $G$. 
However, given that here we take the value  
$\kappa=0.03$, we will ignore this negligible rescaling and 
assume $G$ is the standard value measured by experiment.

Analytical results exist for the above equation in the case of two point 
masses~\cite{Bekenstein:2006fi}, But these are valid only in certain regimes.
The ``deep-MOND'' solution of Ref.~\cite{Bekenstein:2006fi} is valid 
close to saddle, while the ``quasi-Newtonian'' solution is valid for 
large distances. By tackling the problem using numerical techniques we can 
present results accurate even in the gap between these two regimes. This is 
particularly important since this gap is in fact the region of most practical 
importance for probes like LISA Pathfinder. 

Furthermore, since the Earth-Sun saddle point lies within the orbit of the
Moon, we must concern ourselves with the effect of the Moon on 
the LISA Pathfinder measurements, and hence consider the more 
complex three-body problem. Our numerical approach enables us to do just that,
and also to explore the possible advantages of probing
the hitherto unexplored lunar saddle point. 

This article is laid out as follows. In Section~\ref{sec:method} 
we present the algorithm employed in our numerical code. Then in 
Section~\ref{sec:twoBody} we test the code by comparing its results with 
the analytical solutions previously developed for the two-body 
problem in~\cite{Bekenstein:2006fi}. Aware of the virtues and limitations
of the code in Section~\ref{threeBody} 
we explore the solution for the full three-body problem, in 
a number of configurations, and considering both 
the Earth-Sun and Moon-Sun saddles. Specific recommendations for the
LISA Pathfinder mission are made in Section~\ref{sec:LISA} and more
general considerations are included in the last Section. In Appendix~\ref{sec:numerics}
we present details of our numerical algorithm.


\section{Numerical method for solving the non-relativistic equations}
\label{sec:method}

\begin{figure}
\resizebox{0.7\columnwidth}{!}{\includegraphics{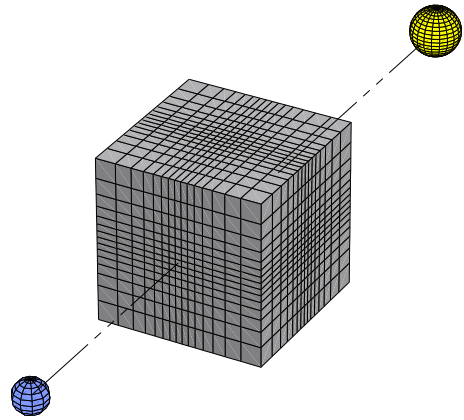}}
\caption{\label{fig:grid}An illustration of the lattice location when calculating for the Earth-Sun saddle region. The calculation is performed for the a cuboid surrounding the saddle region, but not enclosing any of the gravitating bodies. Also shown is the type of non-uniform coordinates that we consider in order to boost the resolution in the central regions of the lattice, i.e. near the saddle point. For example, the $y$-resolution is a function of the $y$ coordinate only. Not to scale.}
\end{figure}

In this section we present an overview of our numerical approach for the 
solution of Eq. \ref{eqn:TeVeS} (see Appendix~\ref{sec:numerics} for full details). 
Most importantly this involves solving numerically only for the region 
immediately surrounding the saddle point, as shown in \Fig{grid}. 
This region contains no gravitational sources, so that \Eq{TeVeS} 
becomes homogeneous\footnote{This of course neglects the very small density present in interplanetary space and any effects of the LISA Pathfinder measurement probe.}. The gravitating sources make their presence felt by the 
boundary conditions on our gridded volume. So long as the boundary 
is sufficiently far from the saddle point, so that $1-\mu\ll1$, then 
to a very good approximation, $\PHI$ on the boundary is merely 
$\PHI=\frac{\kappa}{4\pi}\PhiN$, i.e. a rescaled version of the Newtonian potential. 
We also use the rescaled Newtonian solution inside the box
as the initial configuration supplied to our numerical relaxation algorithm.

It turns out that 
solving \Eq{TeVeS} numerically in terms of $\PHI$ is sub-optimal since it 
is only $\vect{\Del}\PHI$ that appears in this equation. However, 
solving in terms of either $\vect{g}=-\vect{\Del}\PHI$ 
or $\vect{u}=-(4\pi\mu/\kappa)\vect{\Del}\PHI$ results in there being 
a further equation added to the system in order to specify 
the curl-free nature of $\vect{g}$ or $\vect{u}/\mu$. 
Here we will use $\vect{g}$, and so obtain:
\begin{eqnarray}
\label{eqn:laplace}
\vect{\Del}\cdot\mu \vect{g} = 0\\
\label{eqn:curl}
\vect{\Del}\times\vect{g} = 0,
\end{eqnarray}
(written in terms of $\vect{u}$, the $\mu$ appears instead in the second equation).
Fortunately we can bypass the extra work involved in dealing with the
extra equation by choosing a 
relaxation algorithm for \Eq{laplace} that preserves the curl 
of $\vect{g}$ as it proceeds. Since we use a  Newtonian initial
configuration, with zero curl, we obtain a solution to \Eq{curl} at 
no additional computational cost. We represent the field $\vect{g}$ 
on a non-uniform lattice of discrete points, as shown in \Fig{grid}. 
Therefore, as we explain in Appendix~\ref{sec:numerics}, in reality 
we obtain a solution to a finite difference version of \Eq{laplace}. 
The algorithm preserves the finite difference version of 
$\vect{\Del}\times\vect{g}$ exactly (i.e. to floating-point accuracy in real life). 

Interestingly, this approach is the reverse of that used by 
Milgrom~\cite{Milgrom:1986ib}, who solved for $\vect{u}$ 
using the Newtonian solution to obtain $\vect{\Del}\cdot\vect{u}=0$ initially. 
The field $\vect{u}$ was then relaxed to a solution in which 
$\vect{u}/\mu$ has zero curl whilst not disturbing the divergence.
However Milgrom considered only 2D problems, for which there is just one curl equation. In 3D there are three curl
equations and it is beneficial to instead obtain the solution to the curl equations 
for free; hence our chosen algorithm. 


\section{The two-body problem}
\label{sec:twoBody}

We first consider a two-body setup, such as the Earth-Sun system in the 
absence of the Moon or any other perturbations. As we shall show later, 
the Moon only perturbs slightly the results for the Earth-Sun saddle 
point and hence this simpler problem provides great insight for the
three-body situation to be explored in the next section. Furthermore
there are analytical solutions for the two-body problem, so it
presents a natural arena for checking our numerical algorithm.
Conversely our numerical results will fix a poorly constrained 
parameter which appears in one of the analytical solutions.

In Ref.~\cite{Bekenstein:2006fi} solutions in two different regimes 
were found for the MONDian two-body problem. The first is valid for 
regions far from the saddle point and was labelled the ``quasi-Newtonian'' 
solution; while the second is valid closer to the saddle and was 
dubbed the ``deep-MOND'' regime. Unfortunately from a practical perspective, it would be difficult to fly a spacecraft very close to the saddle point, where the deep-MOND solution is accurate, due to navigational difficulties but also because, for example, the Earth-Sun saddle point moves as the Moon orbits (see next section) and is perturbed to a lesser extent by other solar system objects and by the Galaxy. In contrast, for large distances from the saddle, where the quasi-Newtonian solution is accurate, the MONDian effects are simply too small. 
Hence it is in fact intermediate distances, between the two regimes, 
which are of most practical interest and for which our numerical 
results have most value.

In this section, unless otherwise specified, 
we will follow Ref.~\cite{Bekenstein:2006fi} and approximate the 
Newtonian acceleration (providing the MOND boundary conditions) by:
\begin{equation}
\label{eqn:linearApprox}
\vect{g}^{\mathrm{N}} = A \left(x\vect{i} - \frac{1}{2}y\vect{j} - \frac{1}{2}z\vect{k} \right).
\end{equation}
where $A$ is a constant (related to masses of the Sun and the Earth and their separation;
see~\cite{Bekenstein:2006fi}), the problem is symmetric about the $x$-axis (connecting Sun and Earth) and the saddle is at (0,0,0). This is a valid approximation in the region near the saddle point and is used in deriving the 
analytical solutions. We shall use it for our numerical work in this Section
(but not in the one that follows) in order to perform a fair comparison 
with the analytical results. 

We take the function $\mu(\kappa |\vect{\Del}\PHI| / a_0)$ to have the 
form implied by:
\begin{equation}
\label{eqn:mu}
\frac{\mu}{\sqrt{1-\mu^4}} 
=
\frac{\kappa}{4\pi a_0} |\vect{\Del} \PHI|.
\end{equation}
This is merely for illustration purposes: it was the form 
chosen in Ref.~\cite{Bekenstein:2006fi} since it eases 
analytical progress (with our choice of $\kappa=0.03$ also following that article). 
For the idealized case of spherically symmetric matter distributions this $\mu$ can be converted into the ratio of Newtonian to (total) MOND accelerations $\tilde{\mu}$, which is often discussed in galactic astronomy (see eg. Ref.~\cite{binney,yusaf2,kostasrev,Cioti,Tiret}). However, galactic observations are of limited use in guiding our choice for $\mu$ since large uncertainties exist in the matter distribution within galaxies, which in practice will not be spherical, and a large extrapolation in acceleration scale is required to obtain predictions for $\tilde{\mu}$ in the regime of interest here. The consideration of alternative $\mu$ functions and the conversion of a null result from a pass of LISA Pathfinder through a saddle region into a constraint on $\mu$ is left for future work~\cite{future}.
 


\subsection{Analytical solutions}

The analytical solutions are most easily expressed in terms of a dimensionless version of $\vect{u}=-(4\pi\mu/\kappa)\vect{\Del}\PHI$, given by:
\begin{equation}
\vect{U}=\frac{\kappa^{2}}{16\pi ^2 a_0} \vect{u}=-\frac{\kappa}{4\pi a_0}
\mu \vect{\Del}\PHI.
\end{equation}
This may be also related to the acceleration via:
\be 
\label{eqn:g2U}
\vect{g} = -\vect{\Del} \PHI
={4\pi a_0 \over \kappa}(1+U^2)^{1/4}{\mathbf{U} \over U^{1/2}}\,, 
\label{gradphi} 
\ee
for the chosen $\mu$ form. The field $\vect{U}$ must be divergence free in vacuum in order to satisfy \Eq{TeVeS}, while in order to keep $\phi$ curl-free for this $\mu$ function, $\vect{U}$ must satisfy:
\begin{equation}\label{eqn:Ucurl}
4(1+U^2)U^2 \; \vect{\Del}\times\vect{U} + \vect{U}\times\Del U^2 = 0.
\end{equation}
The deep-MOND regime is characterized by $U\ll 1$, when the $(1+U^2)$ term can be set to 1, whereas the quasi-Newtonian limit is characterized by $U\gg 1$. A transition region separates the two, located near the ellipsoid:
\begin{equation}
\label{eqn:r0ellipse} 
x^2 + \frac{y^2+z^2}{4}  = r_0^2, 
\end{equation}
where $r_{0}$ is the key length-scale of problem, given by:
\begin{equation}
\label{eqn:r0exp}
r_0 = \frac{16\pi^2 a_0}{ \kappa^2 A}.
\end{equation}
We have that $r_0\approx381\unit{km}$ for the Earth-Sun saddle point (ignoring all other gravitating bodies).

\subsubsection{Quasi-Newtonian solution}

In the quasi-Newtonian regime, well outside ellipsoid (\ref{eqn:r0ellipse}), 
the solution can be conveniently decomposed as:
\begin{equation}
\vect{U}={r\over r_0}{\mathbf N}(\psi) + {r_0\over r}{\mathbf B}(\psi)
\end{equation}
where ($r$,$\psi$,$\theta$) is a spherical coordinate system centred on the saddle point with the two gravitating bodies located at $\psi=0$ and $\psi=\pi$. The first term is just the Newtonian acceleration multiplied by $\kappa^{2}/16\pi^2{a_0}$, and the second ``magnetic'' term has finite curl and is sub-dominant. The angular profiles are  given by:
\bea {\mathbf N}(\psi)&\equiv & N_r {\mathbf  e}_r
+N_\psi{\mathbf e}_\psi \label{newt1}
\\
N_r&=&{\scriptstyle 1\over\scriptstyle 4}[1+3\cos(2\psi)]\\
N_\psi&=&-{\scriptstyle 3\over\scriptstyle 4}\sin(2\psi).
\label{newt2}
\eea
for the Newtonian component, and by
\bea
{\mathbf B}(\psi)&=&B_r(\psi)\ve_r+B_\psi(\psi)\ve_\psi\\
B_r&=&{2\over 5+3\cos2\psi}+{\pi\over 3\sqrt 3},\\
B_\psi&=&{\tan^{-1}(\sqrt 3 -2 \tan{\psi\over 2})+
\tan^{-1}(\sqrt 3 +2 \tan{\psi\over 2})\over \sqrt 3 \sin\psi}\nonumber\\
&& -{\pi\over 3\sqrt 3}\frac{\cos\psi+1}{\sin\psi}.
\eea
for the magnetic component (these formulae are all derived in~\cite{Bekenstein:2006fi}). 
Note that it is only in this regime that the Newtonian-like and curl
components can be split in this manner. 

\subsubsection{Deep-MOND solution}

Inside the ellipsoid (\ref{eqn:r0ellipse}) the curl contribution becomes 
essential, and cannot be disentangled from the full field. 
A semi-analytical solution was found in Ref. \cite{Bekenstein:2006fi} to be:
\begin{equation}
 \mathbf{U}=C\Big({r\over r_0}\Big)^{\alpha-2}(F(\psi)\,
\mathbf{ e}_r+G(\psi)\,\mathbf{ e}_\psi)
\label{Ufinal}
\end{equation}
with $\alpha\approx 3.528$, and angular profile:
\bea \nonumber
F(\psi)&=&0.2442+ 0.7246 \cos(2\psi) + 0.0472 \cos(4\psi) + \ldots,
\\
G(\psi)&=&-0.8334 \sin(2\psi) - 0.0368 \sin(4\psi) + \ldots ,
\eea
which is almost identical to the Newtonian profile ($F\approx N_r$
and $G\approx N_\psi$). These formulae were derived in~\cite{Bekenstein:2006fi}.

The normalization $C$ is set by the boundary 
conditions and matching the two solutions suggests $C\approx 1$.
Here we will determine $C$ by comparison with our numerical results.


\subsection{Numerical results and comparison to analytical solutions}

\begin{figure}
\resizebox{\columnwidth}{!}{\includegraphics{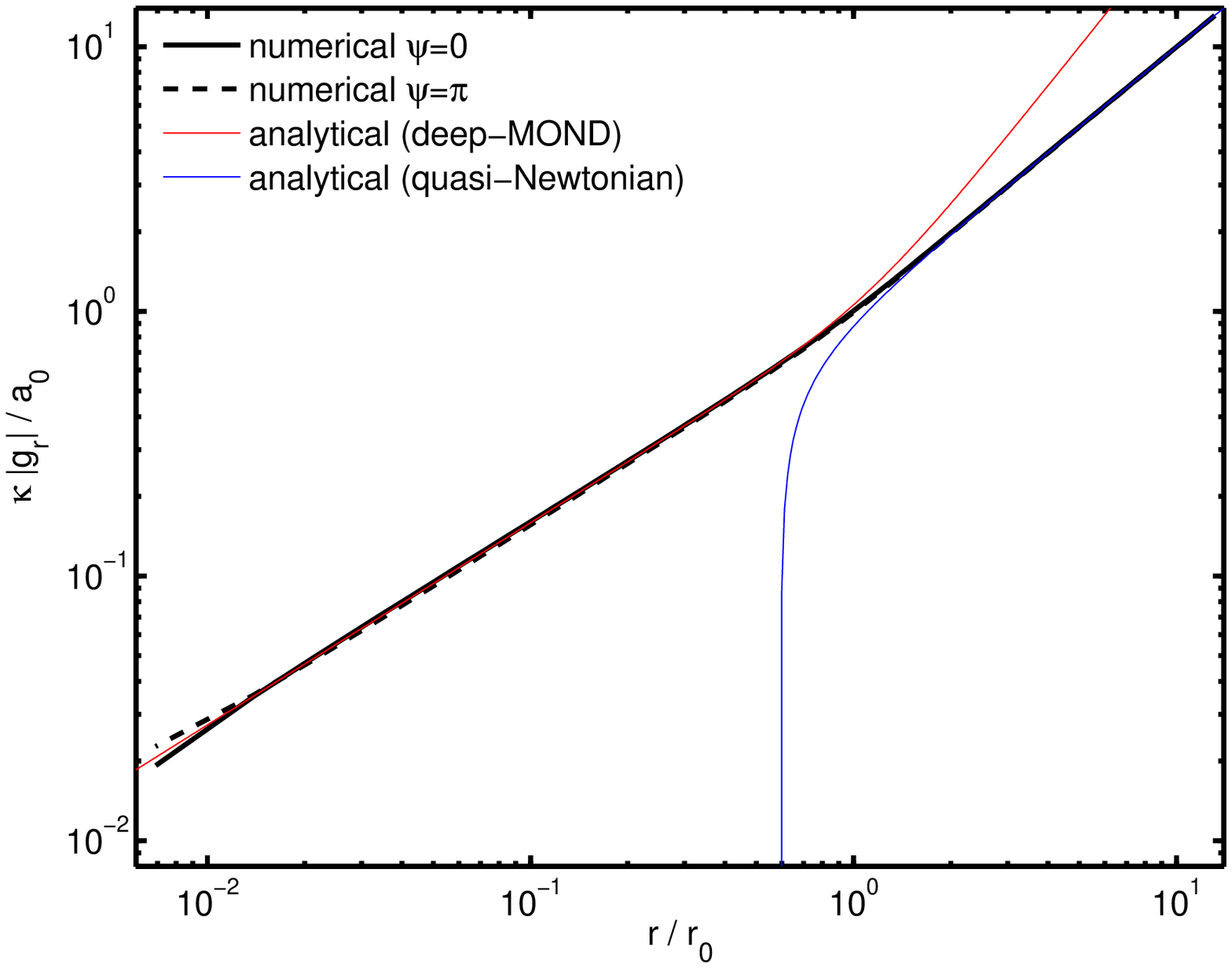}}\\
\resizebox{\columnwidth}{!}{\includegraphics{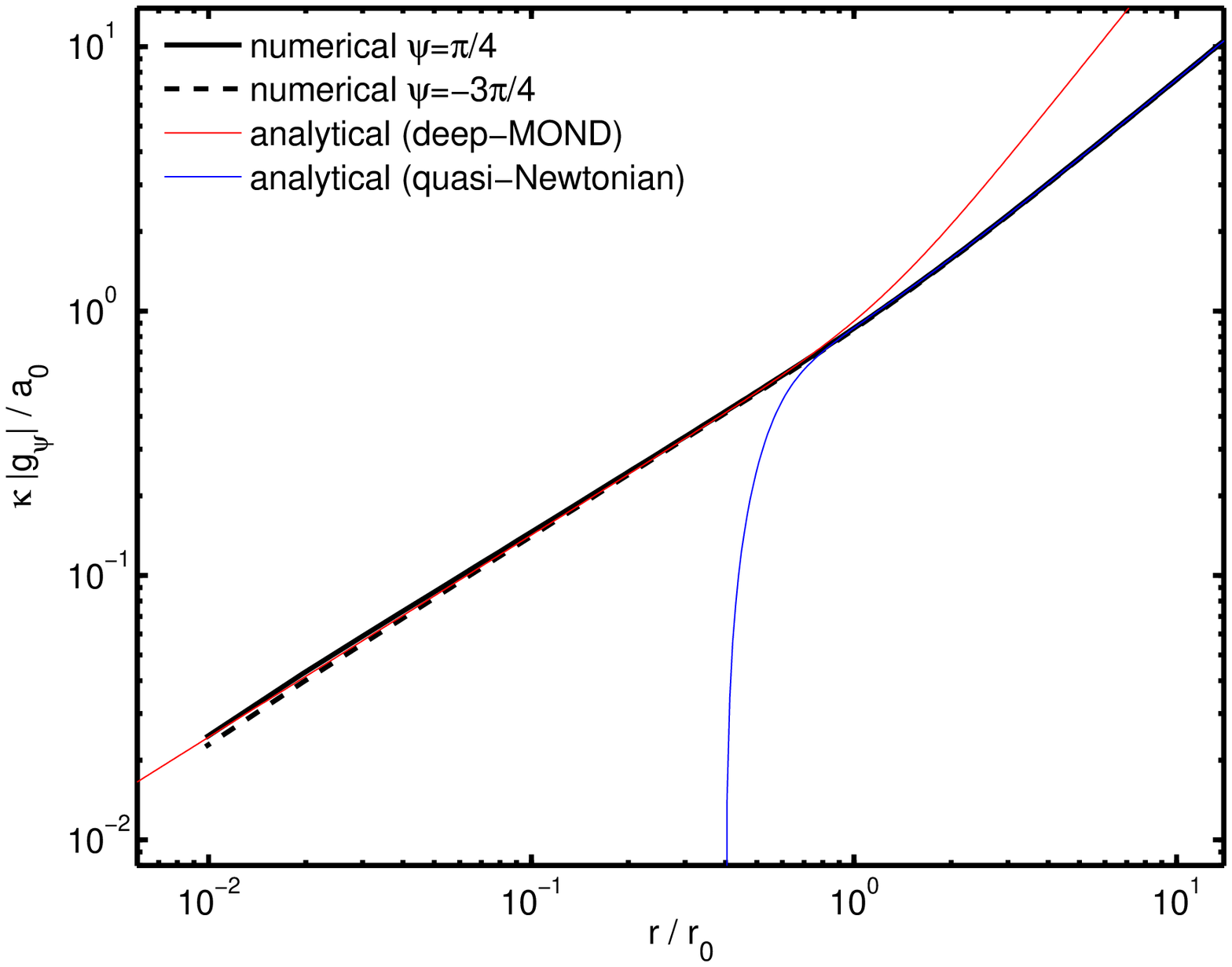}}\\
\resizebox{\columnwidth}{!}{\includegraphics{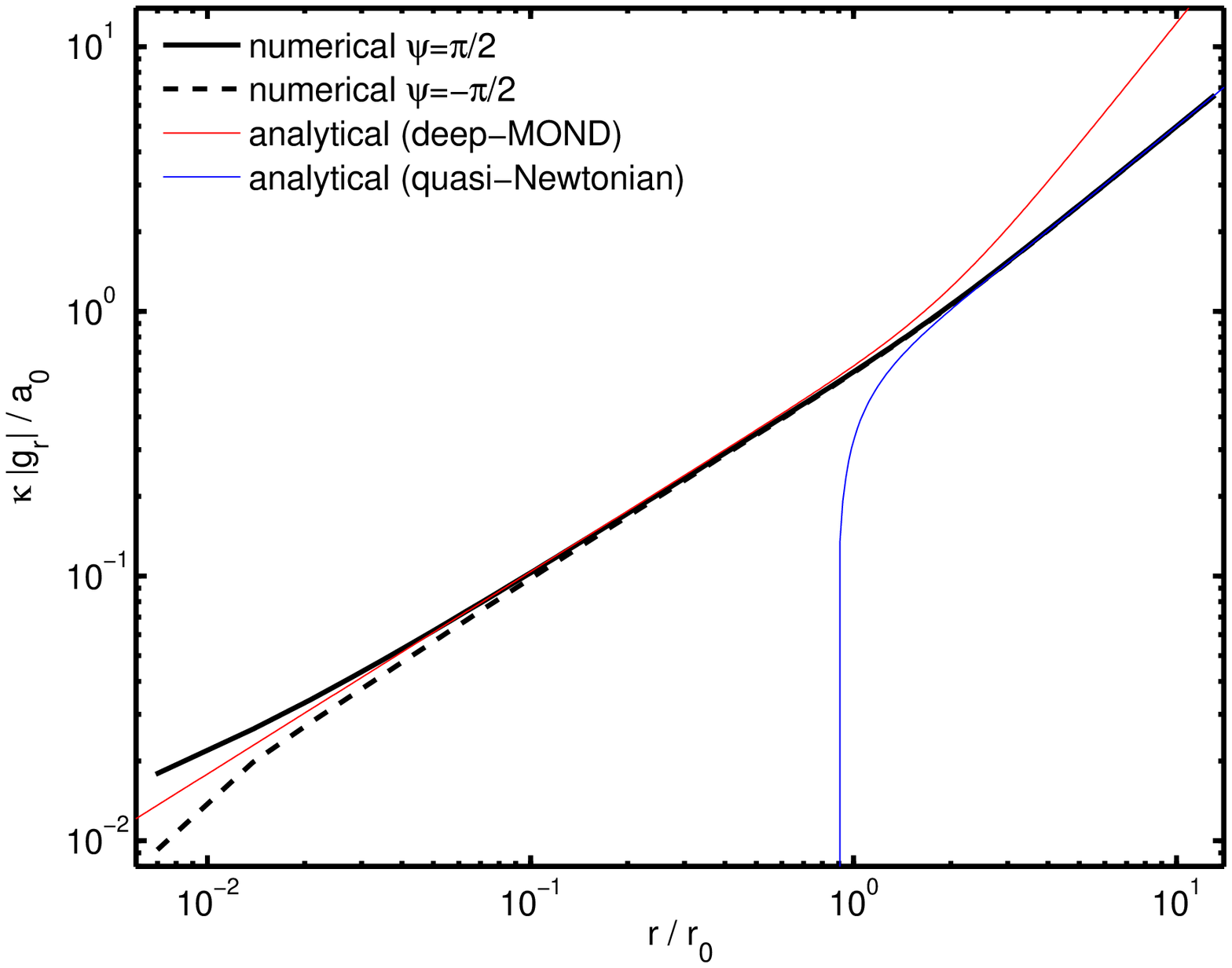}}
\caption{\label{fig:analyComp}A comparison between the numerical and analytical results for components of $\vect{g}=-\vect{\Del}\phi$ in the two-body Earth-Sun case. Results are plotted as function of $r$ for three pairs of $\psi$ values: $0$ and $\pi$ (top), $\pi/4$ and $3\pi/4$ (middle) and $\pm\pi$ (bottom), with symmetry relating the $\vect{g}$-component values within in each pair up to the sign in the analytical case. In the numerical case this is also approximately true, except for very low $r$ where the discretization asymmetry prevents it (this serves as an estimate of the discretization errors). Note that $g_\psi$ is zero analytically for $\psi=0$ and $\pi/2$ at all $r$, and that $g_r$ tends to zero for $\psi=\pi/4$ at in the large $r$ limit. We have used $C=0.839$ for the deep-MOND analytical results in this figure.}
\end{figure}

We have first applied our numerical algorithm to the case where the Newtonian acceleration on the lattice boundary obeys the linear approximation to the two-body problem given by \Eq{linearApprox}, exactly as in the above analytical results. This will allow a fair comparison between the two approaches. Considering the Earth-Sun saddle ($r_0\approx 381\unit{km}$) and using a $257^{3}$ lattice of physical extent $10\;000\unit{km}$ and central resolution $\approx2.6\unit{km}$, we find results for $\vect{g}$ as illustrated in \Fig{analyComp}. These can be seen to yield a good match to the two analytical solutions within their respective domains and provides the appropriate interpolation between them. 

Due to discretization asymmetries, the position of $\vect{g}=\vect{0}$ is not precisely at the central site of the lattice. The lowest value of $|\vect{g}|$ is instead found on a nearby site, but this is just a few kilometres away and represents good accuracy considering the $10\;000\unit{km}$ box size. Before performing the comparison against the analytical results we first translate our numerical solution to take this offset into account, which can be important at very small $r$ values. Without this, the gap between the thick solid and dashed lines in \Fig{analyComp} would have been more noticeable.

The numerical results enable us to determine the previously poorly constrained $C$ parameter appearing in the deep-MOND analytical solution. To determine this we first converted $\vect{g}$ into $\vect{U}$ via \Eq{g2U} and then determined the ratio of the numerical results to the $C=1$ analytical values for all lattice sites within bounds of $r/r_0 = 0.05 \rightarrow 0.5$, chosen to be comfortably in the deep-MOND regime. We find a ratio of \mbox{$C=0.839\pm0.016$} at each site, the central value from which we have used for all comparisons against the deep-MOND solution.

However, $\vect{g}$ is not the key measurable quantity. We are instead interested in the tidal stresses, to which LISA Pathfinder is sensitive. When calculating the observable stresses we must subtract from $\vect{g}$ the unobservable rescaled Newtonian contribution (see discussion on the rescaling of $G$ in the Introduction). Hence we introduce the notation: 
\begin{equation}
S_{ij} = - \frac{\partial^2 \PHI}{\partial x_{i} \partial x_{j}} + \frac{\kappa}{4\pi} \frac{\partial^2 \Phi^{N}}{\partial x_{i} \partial x_{j}},
\end{equation}
for the observable MOND stress tensor, where $x_{i}=x$, $y$, or $z$. Under the linear approximation to the Newtonian 
acceleration field (Eq. \ref{eqn:linearApprox}), the Newtonian stress tensor is just \mbox{$S_{ij}^{N}=A\;\textrm{diag}(1,-1/2,-1/2)$} and hence we must simply subtract a constant tensor from the raw MOND results. 

\begin{figure}
\resizebox{\columnwidth}{!}{\includegraphics{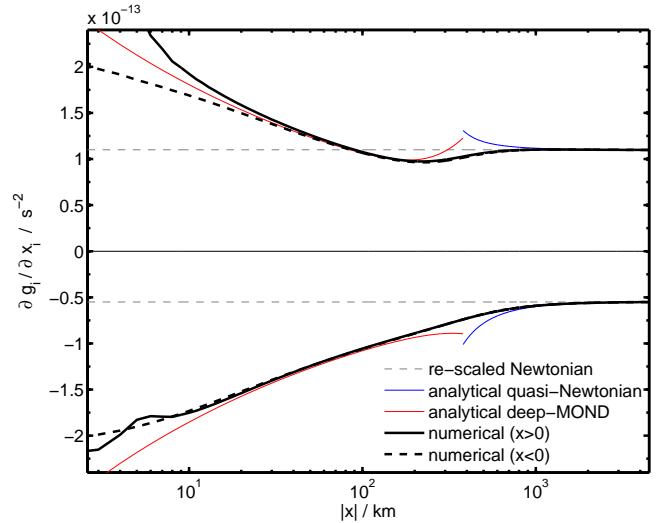}}\\
\caption{\label{fig:analyCompStress}The full contribution to the two-body tidal stresses from the MOND field and the re-scaled Newtonian contribution that must be subtracted from them in order to yield the observable component. The positive results are for $\partial g_x/\partial x$ while the negative ones are for $\partial g_y/\partial y$. Results are shown for the line $y=z=0$, for which the analytical results shown are symmetric to the interchange $x\rightarrow-x$. This symmetry is not quite realized in numerical case, as can be seen by the small differences between the dashed ($x>0$) and solid lines ($x<0$) for the numerical solution. The numerical stresses become unreliable for $|x|<20\unit{km}$.} 
\end{figure}

\begin{figure}
\resizebox{\columnwidth}{!}{\includegraphics{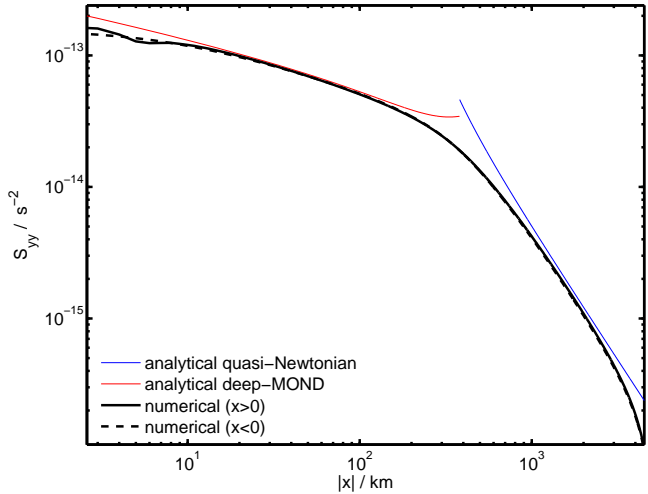}}\\
\caption{\label{fig:analyCompStressLog}The observable contribution to the $\partial g_y/\partial y$ tidal stress from the MOND field for line $y=z=0$. It can be seen that the the numerical results perform well even after the subtraction of the unobservable re-scaled Newtonian contribution, which heavily dominates the stresses in the quasi-Newtonian regime. However, near the edge of the simulation volume the stress is attenuated due to the smooth transition of $\phi$ to meet the rescaled-Newtonian boundary conditions.}
\end{figure}

We illustrate the nature of this subtraction in 
\Fig{analyCompStress}
and
\Fig{analyCompStressLog}, where it can be seen that the constant unobservable contribution dominates the full stress in the quasi-Newtonian regime (for diagonal elements of the tensor). As a result of this any inaccuracies in our numerical results become more important relative to the size of the signal, once the subtraction occurs. Furthermore, when calculating the stress there are finite differencing errors, which become increasingly important at small $|x|$. This is evidenced by the differences between the dashed ($x<0$) and solid ($x>0$) lines for the numerical results, and their difference relative to the deep-MOND analytical result. Hence neither regime is trivial computationally. However, as already noted, for LISA Pathfinder we are fortunately interested in the intermediate regime where the reliability of results is greatest.

\begin{figure}
\resizebox{\columnwidth}{!}{\includegraphics{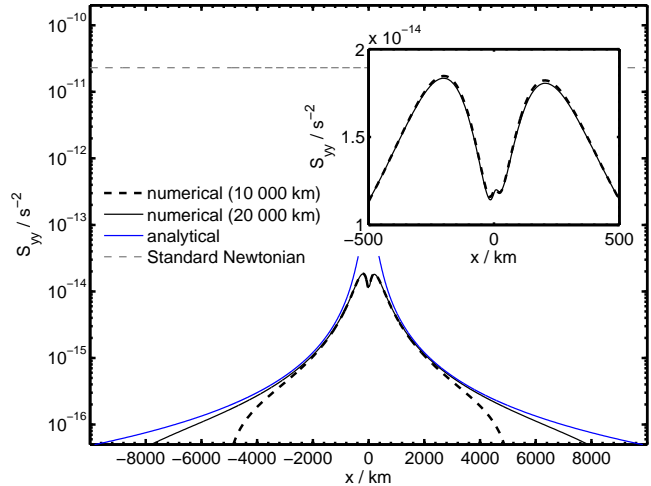}}
\caption{\label{fig:2bodyLinUncert}The two-body MOND stress signal $S_{yy}$ along the line $y=100\unit{km}$ for two simulation box sizes ($10\;000\unit{km}$ and $20\;000\unit{km}$) with fixed lattice size $257^{3}$. The zoomed region with linear $y$-axis shows the minor differences between these two sets of results near the point on this trajectory that is closest to the saddle, in addition to minor asymmetries which are due to discretization errors and more significant for the larger box. The log-scale graph shows the tails of the signal, which are improved in the larger box.}
\end{figure}

When a real spacecraft performs a saddle fly-by, it will not pass precisely through the saddle point. We discuss the likely impact parameter and trajectory for LISA Pathfinder in \Sec{LISA}, but for illustrative purposes we consider here the form of $S_{yy}$ and the numerical uncertainties in it for a trajectory along the line $y=100\unit{km}$. While the normal Newtonian signal is much larger than that from $\phi$, the former is simply constant (though it would slowly drift without the linear approximation used here). The MOND signal, on the other hand, provides a distinctive variation as the probe passes by the saddle. The form is stable against numerical errors, although minor asymmetries can be seen near the point of closest passage. These are slightly more significant when the lattice spacing is increased by enlarging the box side to $20\;000\unit{km}$ (from the $10\;000\unit{km}$ used thus far). However, the larger box provides greater accuracy in the less important tails of the signal (see \Fig{2bodyLinUncert}).

\begin{figure}
\resizebox{\columnwidth}{!}{\includegraphics{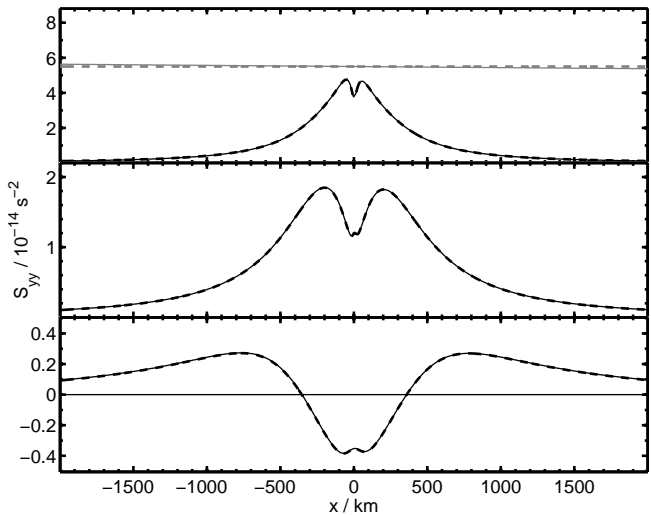}}
\caption{\label{fig:2bodyStress}The MOND stress signal $S_{yy}$ along the lines $y=25$, $100$ and $400 \unit{km}$ (top to bottom) for the full two-body Earth-Sun case (thick black dashed) and when using the linear approximation to the Newtonian acceleration (black solid). In the upper panel the rescaled Newtonian stress is shown (grey) for the $y=25\unit{km}$ case.} 
\end{figure}

To close this Section we 
present results using the full two-body Newtonian acceleration field for the Earth-Sun system, rather than the linear approximation (\Eq{linearApprox}). As can be seen in \Fig{2bodyStress}, the effect of the approximation is insignificant. \Fig{2bodyStress} also shows the change in the form of the signal for three different impact parameters: $y=25$, $100$ and $400\unit{km}$. We see that the signal broadens with increasing $y$, but more importantly, its amplitude decreases. 


\section{The three-body problem}
\label{threeBody}

In this section we add a third body to the problem and consider the perturbing effect of the Moon on the Earth-Sun saddle. We also present results for the lunar saddle point (which is intrinsically a three-body problem, as we shall see).

\begin{figure}
\resizebox{\columnwidth}{!}{\includegraphics{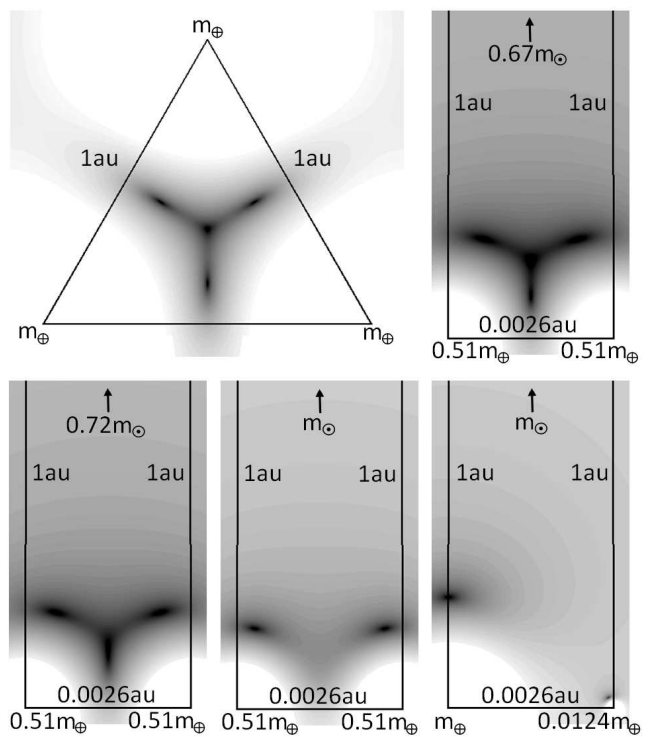}}
\caption{\label{fig:saddleCount}Maps of the Newtonian $|\Del \PhiN|$ for different configurations of 3 point masses, showing differing numbers of zero solutions. Black regions denote low $|\Del \PhiN|$ values while white denotes high values. The upper-left map is for the case of equal masses arranged in an equilateral formation, which yields 3 saddle points plus a central maximum in $\PhiN$. The second frame shows the same number of solutions, but is for a configuration much closer to the Sun-Earth-Moon system, albeit that the Earth's mass is distributed evenly between the lighter two bodies and the heavier body is slightly lighter than the Sun. Mass is added to the heavier body in the third frame, causing the lower saddle and central maximum to approach each other. Adding yet further mass causes them to meet and disappear, leaving only two saddle points by the time the solar mass is reached, as in the fourth frame. Giving the lightest two bodies masses corresponding to the Earth and the Moon, then maintains this number of saddle points but, as can be seen in the final frame, the lunar saddle point is surrounded by only a small region of low acceleration and it is also significantly tilted towards Earth.}
\end{figure}

We must first comment that the ``realistic Solar system'', as described in Ref.~\cite{Bekenstein:2006fi}, 
is in fact not realistic at all, and numerical work is needed to provide even 
qualitatively correct directions for a space mission. The argument in Ref.~\cite{Bekenstein:2006fi} regarding the Earth-Moon saddle, for example, is entirely flawed because of its use of the linear approximation for the Newtonian field for the Earth-Moon system while ignoring the gravitational effects of Sun (cf. Eq.~(72) in that paper). 
As it happens, the perturbation induced by the Sun is too large for this approximation to be valid and in fact there is no true Earth-Moon saddle point. Instead as explained in Fig.~\ref{fig:saddleCount}, the Sun-Earth-Moon system presents only two saddle points: the Sun-Earth saddle, which is more or less as the two-body analysis suggests, and a single saddle point close to the Moon, which is intrinsically a three-body problem. But even for the latter, we shall empirically find a recipe for adapting the two-body results to the real situation.

\subsection{The Sun-Earth saddle}

For the Sun-Earth saddle, the effect of the third body is not drastic, because the lunar mass is approximately $1/81$th that of Earth and even at new Moon, the Earth is only about twice as far away from the saddle point. 

\begin{figure}
\resizebox{\columnwidth}{!}{\includegraphics{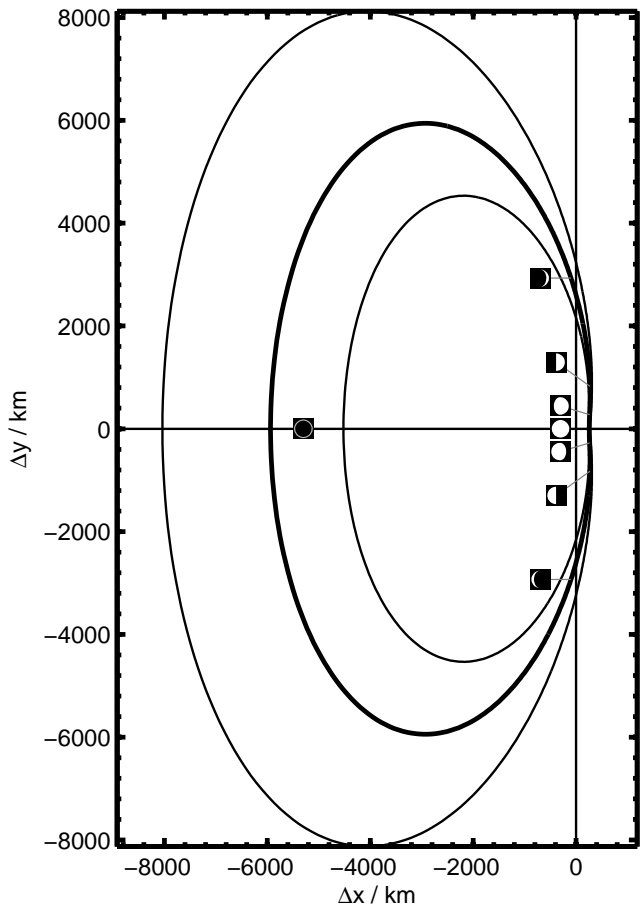}}
\caption{\label{fig:EarthSunSaddlePerturb}The perturbation in the position of the Earth-Sun saddle due to the presence of the Moon when the lunar orbital radius is fixed at its semi-major axis (thick), with addition plots for apogee (outer) and perigee (inner) to highlight the effect of orbital ellipticity. Negative $\Delta x$ values denote that the saddle moves closer to Earth, which would lie to the left in this figure, while the Sun would lie to the right. The saddle is only slightly perturbed at Full Moon, moving slightly towards the Sun, but as new Moon approaches the perturbation becomes large and saddle moves quickly, covering the left half of the quasi-ellipse in approximately 3 days, with new Moon being the mid-point of this period. The perturbation at new Moon is enhanced if new Moon coincides with the lunar perigee but attenuated if at apogee.}
\end{figure}

We will firstly consider the location of the saddle, which is of great practical importance when planning a spacecraft fly-by. It must be considered that the length scale on which the MOND signal is significant, $r_{0}$, is approximately $381\unit{km}$ (see Eq.~\ref{eqn:r0ellipse}), so the saddle has to be located to a precision better than this. The saddle point is approximately \mbox{$258\;800\unit{km}$} from Earth (around two-thirds to the lunar orbital radius), and therefore even a small perturbation is potentially important. As shown in Fig.~\ref{fig:EarthSunSaddlePerturb}, at full Moon the saddle shifts about $250\unit{km}$ towards the Sun. The magnitude of the shift remains at approximately this level until the crescent phase approaches, when the Moon begins to approach the saddle. The saddle then starts to experience a large  perturbation and, in the 3 day period surrounding new Moon, it quickly moves over the left half of the quasi-ellipse indicated in the figure. Note that the effect near the new Moon is significantly enhanced if the Moon is at its perigee during this period. While the perturbation is always less than \mbox{$10\;000\unit{km}$}, and is less than $1\;000\unit{km}$ for most of the lunar cycle, this cannot be ignored if a spacecraft is to be navigated through the saddle region.

While the perturbation caused by the Moon generally breaks the axial symmetry present in the two-body discussion of the previous chapter, during a New or Full Moon the symmetry remains unbroken. Hence we may consider these two extreme cases via the same linear approximation to the Newtonian acceleration, but with differing $A$ values (cf. \Eq{linearApprox}). We can then make the following scaling argument. The conditions on the field $\vect{U}$ are that it must be divergence-free, satisfy \Eq{Ucurl} and match the boundary conditions. The first two of these are unaffected by a rescaling of the spatial coordinates $\vect{r}\rightarrow\alpha\vect{r}$ while the boundary conditions are such that: 
\begin{equation}
\label{eqn:linearApprox1}
\vect{g} \rightarrow \frac{4\pi a_0}{\kappa} \left(\frac{x}{r_0} \vect{i} - \frac{1}{2}\frac{y}{r_0}\vect{j} - \frac{1}{2}\frac{z}{r_0}\vect{k} \right)
\end{equation}
in the linear approximation of \Eq{linearApprox}. Hence in scaled coordinates $\vect{r}/r_0$ there is a single solution for $\vect{U}$ valid for all $A$. Therefore while the form of $S_{ij}$ stresses is unaffected by the change in $r_0$, the spatial extent of the signal is proportional to $r_0$ and the stress magnitude scales inversely with $r_0$. (This is a general argument to be used later in the Moon saddle analysis).

At Full Moon the change in $r_{0}$ is minor, increasing by about $1\unit{km}$ from $381\unit{km}$. However at new Moon, the change is more significant, with $r_{0}$ decreasing to $327\unit{km}$~\footnote{These figures assume the Moon-Earth separation is equal to the semi-major axis for the lunar orbit; for the lunar perigee this value should be decreased by $22\unit{km}$ and for the apogee it should be increased by $15\unit{km}$.}. Hence the Newtonian stresses will be slightly higher at new Moon than in the two-body case, and very slightly lower in the Full Moon case. Considering these changes at a fixed $r/r_0$ value, the MOND $\vect{g}$ will be unaffected, therefore $S_{ij}$ at fixed $r/r_0$ increases in inverse proportion to $r_0$. However, note that at new Moon, the linear approximation is not so robust because the gravity field of the Moon varies on a smaller length scale than that of the Earth. 

\begin{figure}
\resizebox{\columnwidth}{!}{\includegraphics{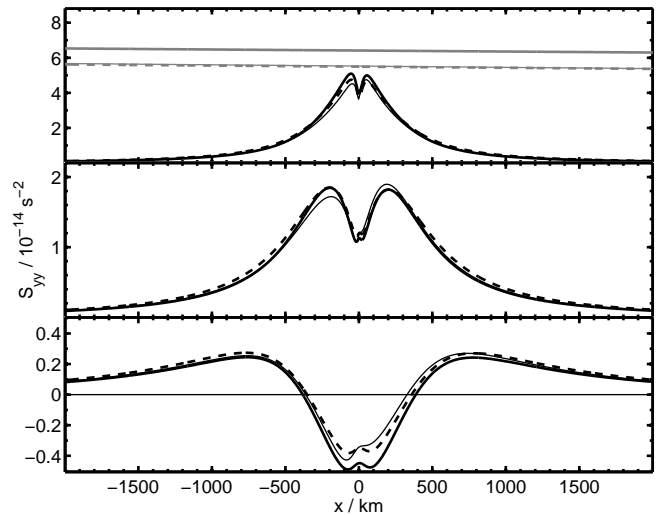}}
\caption{\label{fig:3bodyStress}The MOND stress signal $S_{yy}$ along the lines $y=25$, $100$ and $400 \unit{km}$ (top to bottom) for the three-body Earth-Sun case for different lunar phases: new Moon (thick, black, solid), Full Moon (thick, black, dashed) and when the Moon appears $18^\circ$ away from the Sun towards positive $y$ (thin, black, solid). Also shown in the $y=25\unit{km}$ case are the rescaled Newtonian stresses (grey).} 
\end{figure}

To consider an arbitrary Moon position, and without the approximation to the Newtonian acceleration field, we require numerical methods. The results in \Fig{3bodyStress}, which shows three different lunar phases, indicate that the effect of the Moon on the MOND signal near the Earth-Sun saddle point is minor. At Full Moon the signal is essentially identical to the two-body case, since the Moon is more than twice as far from the saddle point as Earth. The saddle is displaced furthest away from the Earth-Sun line when the angle between the Moon and the Sun (measured from Earth) is approximately $18^\circ$, and even then the line in the figure that corresponds to this asymmetric case still has the same basic form as the two-body signal. Some $x\rightarrow-x$ asymmetry is instilled, but the overall conclusion is that the MOND signal for the Earth-Sun saddle is largely unaffected by the phase of the Moon.


\subsection{The Lunar saddle}

The lunar saddle can potentially provide significantly larger stress signals than the Earth-Sun saddle and is therefore of potential interest experimentally. However, the larger stress signal implies that the MOND region is smaller and so the challenge of navigating a space probe through the saddle region becomes more difficult. 

\subsubsection{Newtonian gravity field}

We must first consider the lunar saddle from a purely Newtonian perspective, since the Newtonian physics sets the saddle location and the boundary conditions for our MOND calculation. It is useful to think of the lunar saddle as a heavily perturbed Moon-Sun saddle, since the Earth is the least important of the three bodies in the saddle vicinity. 

Although the gravitational pull of the Sun is effectively constant on the scale of the lunar orbit, the Moon is approximately $1/81$th of the mass of Earth, and therefore the distance from the unperturbed Moon-Sun saddle to the Moon should be around $1/9$th of the distance from the Earth-Sun saddle to the Earth. Since tidal stresses follow an inverse-cube law then the Sun makes only a minor contribution to the tidal stresses at either saddle, but the Moon causes stresses at the unperturbed Moon-Sun saddle  around $9$ times stronger than the tidal stresses caused by the Earth on the Earth-Sun saddle. This implies (cf. Eq.~\ref{eqn:r0exp}) that the value of $r_0$ for the lunar saddle should be around $9$ times smaller than that for the Moon-Sun saddle, before third-body perturbations are considered.

\begin{figure}
\resizebox{\columnwidth}{!}{\includegraphics{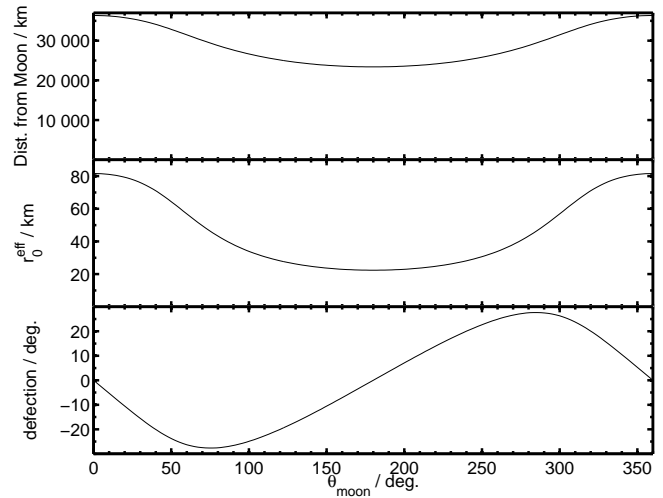}}
\caption{\label{fig:lunarR0}Upper panel: The distance of the lunar saddle from the Moon as a function of the lunar phase, where $\theta_\mathrm{moon}=0$ denotes new Moon and $\theta_\mathrm{moon}=180^\circ$ denotes Full Moon. Middle panel: The effective $r_0$ value for the lunar saddle as a function of lunar phase. Lower panel: The deflection angle of the saddle away from the Moon-Sun line (towards Earth).} 
\end{figure}

\begin{figure}
\resizebox{\columnwidth}{!}{\includegraphics{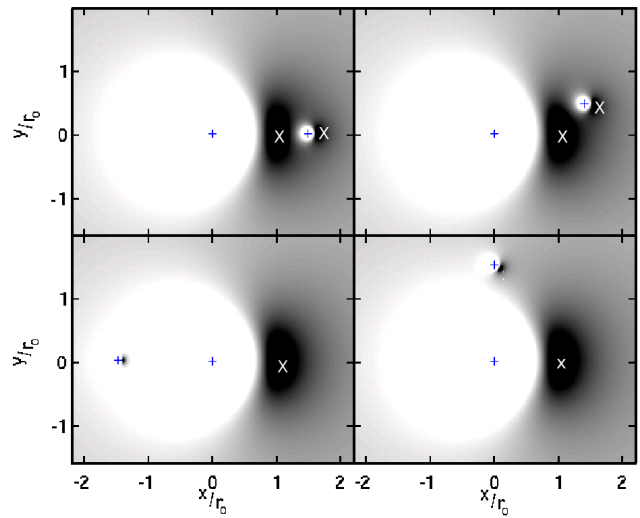}}
\caption{\label{fig:Gcontours}Four maps of the Newtonian acceleration strength $|\vect{g}^{\mathrm{N}}|$ (as before dark means small $|\vect{g}^{\mathrm{N}}|$, and so a saddle, indicated by a X wherever it makes it clearer) in the vicinity of Earth for Sun-Earth-Moon system (in plane $z=0$). Clockwise from top left the angle between the Moon and the Sun, measured at Earth are $0^{\circ}$ (new Moon), $18^{\circ}$, $90^{\circ}$ (First Quarter) and $180^{\circ}$ (Full). The Earth (0,0) and the Moon are shown by crosses. White corresponds to values greater than $80\times10^6 a_{0}$, black indicates values less than $25\times10^6 a_{0}$, with shades of gray indicating intermediate values. Note that the Earth perturbs the Moon-Sun saddle region heavily, both in orientation and size.} 
\end{figure}

However the perturbing effect of the Earth on the lunar saddle is very large, as can be seen in Figures \ref{fig:lunarR0} and \ref{fig:Gcontours}. At new Moon the Earth helps the Moon overcome the pull from the Sun and hence the saddle is further from the Moon than on average and the lunar stress contribution is smaller, yielding a larger saddle region. At Full Moon, the Earth acts with the Sun and therefore the saddle is closer to the Moon, the stresses are greater and the saddle region becomes very small. Additionally, the Earth shifts the lunar saddle significantly away from the Moon-Sun line, by up to $\approx30^\circ$ (measured at the Moon), although of course it lies on this line at full and new Moon (ignoring the possible small displacement of the Moon off the Ecliptic). The lunar saddle is therefore intrinsically a three-body problem.

Importantly, since the Moon dominates the tidal stresses near the saddle, the form of the Newtonian acceleration in the saddle region is still approximately that of the linear approximation form, just as in the Earth-Sun case where the Earth dominates the stress. However, now the effective ``axis of symmetry'' points from the Moon to the saddle location. At once this suggests a recipe for adapting the two-body MONDian solution to the lunar saddle.

\subsubsection{MOND stress signal}

\begin{figure}
\resizebox{\columnwidth}{!}{\includegraphics{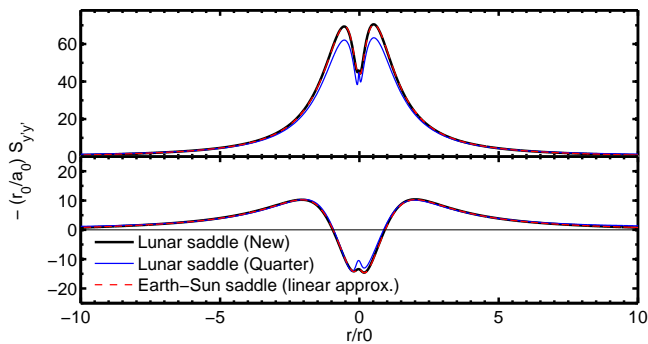}}
\caption{\label{fig:LunarStressMiss}Stress signals for the lunar saddle, compared to the results from the Earth-Sun saddle (in the linear approximation). Lunar results are expressed in a new coordinate system $(x',y',z)$, a rotated version of $(x,y,z)$ such that the $x'$-axis is the line joining the Moon and the saddle, with $y'\approx0.26r_0$ (top) and $1.05r_0$ (bottom). In the Earth-Sun case, the results are for $y=100$ and $y=400\unit{km}$, which correspond to the $0.26r_0$ and $1.05r_0$, making the figure directly comparable with the lower two panels of \Fig{2bodyStress}. At New moon $r_0 \approx 81\unit{km}$ while at Quarter Moon $r_0 \approx 38\unit{km}$.}
\end{figure}

Given the observation that the Newtonian acceleration field near the lunar saddle is roughly linear (as in Eq. \ref{eqn:linearApprox}), it might be expected that the MOND solution would be very similar to the two-body case, except for re-scalings with $r_0$. This is indeed seen in the numerical results displayed in \Fig{LunarStressMiss}, where the new Moon case is indistinguishable from the previous Earth-Sun results and the even the non-axial quarter Moon case is only slightly different, \textit{once appropriate scalings with  $r_0$ are introduced}. For the purposes of this figure we have introduced a new coordinate system $(x',y',z)$, which is a rotated version of $(x,y,z)$ such that the $x'$-axis joins the Moon and the saddle point, which still lies at (0,0,0). We have then chosen $y'\approx0.26r_0$ and $1.05r_0$ lines for which to plot the MOND stress signals, where $r_0\approx81\unit{km}$ at new Moon and $r_0 \approx 38\unit{km}$ at quarter Moon. The two panels in this figure are hence directly comparable with the lower two panels in \Fig{2bodyStress}. However, we have not included results equivalent to the upper panel there ($y'\approx0.066r_0$) since at Quarter Moon this amounts to $y\approx2.5\unit{km}$ which we believe is to too small an impact parameter to consider for LISA Pathfinder (see next section).
%
%

Hence the form of the lunar saddle MONDian signal is essentially the same as for the Earth-Sun saddle; yet it offers larger MOND and Newtonian stresses, but over  a smaller region. In practical terms, however, not much benefit might arise if the navigational constraints dictate a certain minimum likely impact parameter. The MOND signal varies roughly as $1/r$, so for this impact parameter the likely stress level is largely independent of $r_0$. 


\section{Testing MOND with LISA Pathfinder}
\label{sec:LISA}

We now explain more specifically how the results found in this paper
translate into a concrete signal, given the navigational constraints
and instrument noise properties of LISA Pathfinder. We consider one
example here and in a companion paper~\cite{companion} 
perform a more comprehensive study
of the issues highlighted. 


\subsection{Overview of LISA Pathfinder Project}

The LISA Pathfinder (LPF) project~\cite{LISA} is an European Space Agency (ESA) mission designed to test the technology intended to be employed in LISA (Laser Interferometer Space Antenna), a proposed gravity wave observatory. While LISA itself is planned to consist of three spacecraft in a triangular configuration of sidelength $5\times10^{6}\unit{km}$, the Pathfinder mission will consist of two test masses within a single spacecraft and just $35\unit{cm}$ apart. These test masses will follow geodetic motion to exceptional precision
due to protection from solar radiation pressure and other unwanted non-gravitational influences, while the spacecraft itself is designed to minimize its own gravitational impact. An interferometer will study the relative motion of the test masses such that LPF will be an excellent instrument for measuring tidal stresses, such as those predicted in the preceding two sections. The mission is presently in the ``implementation'' phase with its launch expected in 2012. 

The nominal plan for the LPF mission is for it to enter a large-amplitude Lissajous orbit (of near halo-orbit dimensions) around the L1 Lagrange point of the Earth-Sun system, at a distance of $\sim1.5\times10^{6}\unit{km}$ from Earth. Once the planned mission at L1 is complete, it can in principle be extended, by breaking from the Lissajous orbit and passing close to the Earth-Sun saddle point, or alternatively near the lunar saddle. In this section we compare the sensitivity of LPF to the possible MOND signal, given its navigation and measurement capabilities.


\subsection{Spacecraft navigation}

The objective of the LPF mission does not require the spacecraft to have the large thrusters of an interplanetary probe, since the journey out to L1 is made possible by a propulsion module from which LPF then separates. Instead LPF will carry only low-thrust micro-propulsion systems and the ability to navigate the probe near a saddle point is not something that can be taken for granted. 

Fortunately, the Lissajous orbit is unstable and hence LPF can be ejected from the L1 region using a fairly small change in its velocity. For example, a 30-day burn can yield a $\diff v$ of $\sim1 \unit{ms^{-1}}$ and put LPF on a trajectory that brings it close to Earth in a reasonable time-frame, of order one year. The precise orbit depends on the timing of the burn relative to the phase of the Lissajous orbit, the burn magnitude, subsequent correction manoeuvres and any close lunar fly-bys. Without the latter two, approaches near the Earth-Sun saddle in approximately $1\unit{yr}$ tend to have quite large impact parameters $\sim10\; 000\unit{km}$ but closer approaches $<1\,000\unit{km}$ can be achieved using a lunar fly-by. Further correction manoeuvres can aid targeting the saddle region and the likely impact parameter is ultimately determined by the ability to track the spacecraft and apply appropriate $\diff v$ manoeuvres. Due to Earth's gravity the spacecraft will be cruising through the saddles with a speed of the 
order of $\sim 1\unit{kms^{-1}}$. In what follows we will take it that LPF can be directed through a saddle region with velocity $1\unit{kms^{-1}}$ and an impact parameter $\sim 50\unit{km}$. A more complete study of these orbits is presented in~\cite{companion}.


\subsection{Sensitivity to MOND stress signal}

Firstly, it should be noted that the design of LPF limits it to being sensitive only to one of the diagonal elements of the stress tensor, eg. $S_{yy}$ in our notation. Additionally, in order to yield a stable radiation pressure it is desirable to keep fixed the orientation of the (roughly cuboid) probe relative to the Sun, with the side on which the solar panel is mounted facing directly at the Sun. Given the vast distance of the spacecraft from the Sun, the solar panel is therefore effectively aligned normal to the $x$-axis and the spacecraft design then implies that $S_{xx}$ cannot be measured. This leaves the measureable stress as $\cos^{2}(\alpha)S_{yy}+\sin^{2}(\alpha)S_{zz}$, where $\alpha$ specifies the orientation of the test masses relative to the $y$-axis.

Estimates of the sensitivity of LPF as a gradiometer are normally expressed via the power spectral density of the stress signal. This is defined as:
\begin{equation}
P(f) = \frac{2}{T} \left| \int_{-T/2}^{+T/2} \diff t \; S_{ij}(t) \; e^{-2\pi i f t}  \right|^{2}
\end{equation}
where $f$ is the frequency, $t$ is the time and $T$ is the integration period. When integrated over $f$ from $0$ to infinity this yields the mean ``power'' in the signal, i.e. the time-average of $S_{ij}^{2}$, with the factor of $2$ present because negative frequencies are folded in with the positive ones. For the present article we assume that LPF will attain a sensitivity in the square-root of this quantity, the amplitude spectral density, of $1.5\times10^{-14}\unit{s^{-2}/\sqrt{Hz}}$ between $1$ and $10\unit{mHz}$~\cite{LISA}. Beyond this range we will take the sensitivity to degrade as $1/f$ at lower frequencies and as $f^{2}$ at higher frequencies. This yields a good approximation to published expectations~\cite{LISA}.

We simulate gradiometer noise under the assumption of Gaussianity (independent Fourier mode phases) and the above spectrum. We add this to the theoretical stress signal, which is the combination of the standard Newtonian and MOND components, and then apply a simple cleaning algorithm to extract a likely measured MOND signal. Specifically, we approximately remove the Newtonian signal by performing a quadratic fit to the data. In the linear approximation, the Newtonian stress signal is just a constant while the gravitational acceleration near the saddle is so small that we may assume the probe moves with constant velocity and therefore this fit takes into account two further orders, beyond linear, in the spatial Taylor expansion of the Newtonian acceleration. We then apply a top-hat band-pass filter to remove all frequencies outside the range $0.1$ to $10\unit{mHz}$, i.e. keeping those of the approximate order as the MOND signal. 

\begin{figure}
\resizebox{\columnwidth}{!}{\includegraphics{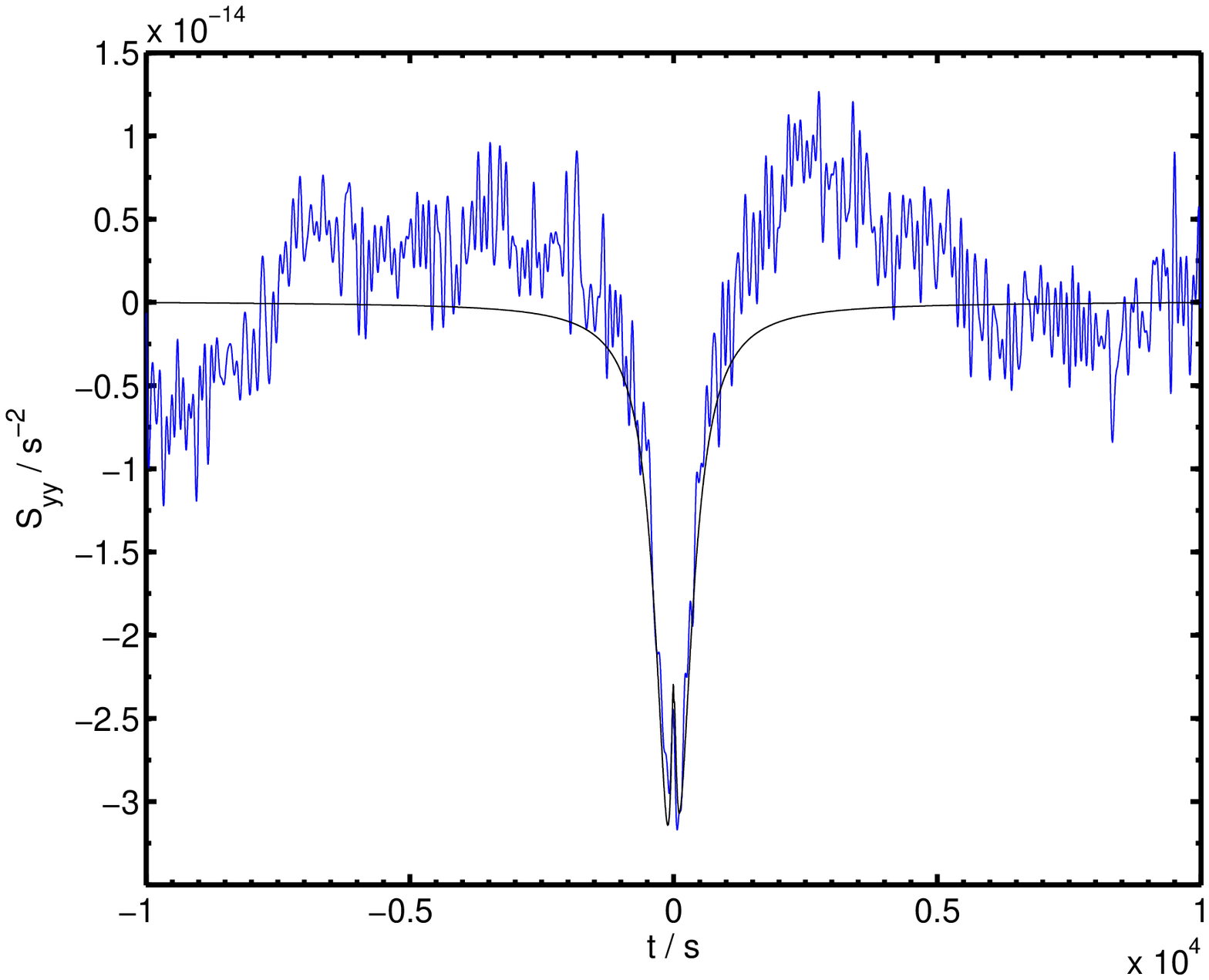}}\\
\resizebox{\columnwidth}{!}{\includegraphics{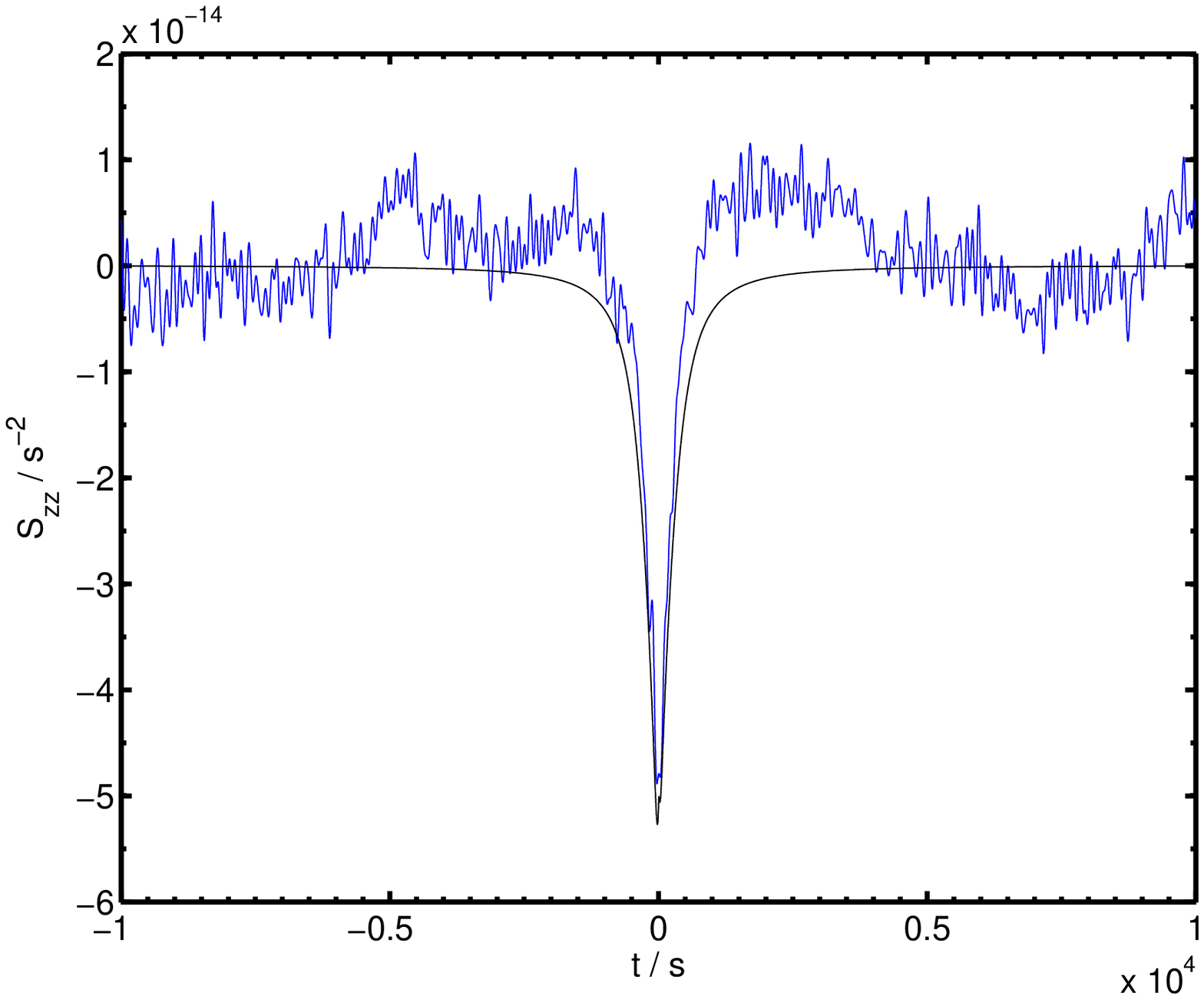}}
\caption{\label{fig:stressNoise}Theoretical and recovered stress signals $S_{yy}$ and $S_{zz}$ for a $y=50\;\unit{km}$, $\dot{x}=1\;\unit{kms^{-1}}$ trajectory, including a realization of gradiometer noise and applying a simple cleaning algorithm (see text). These results are for the three-body Earth-Sun saddle with the lunar phase being full Moon.}
\end{figure}

\begin{figure}
\resizebox{\columnwidth}{!}{\includegraphics{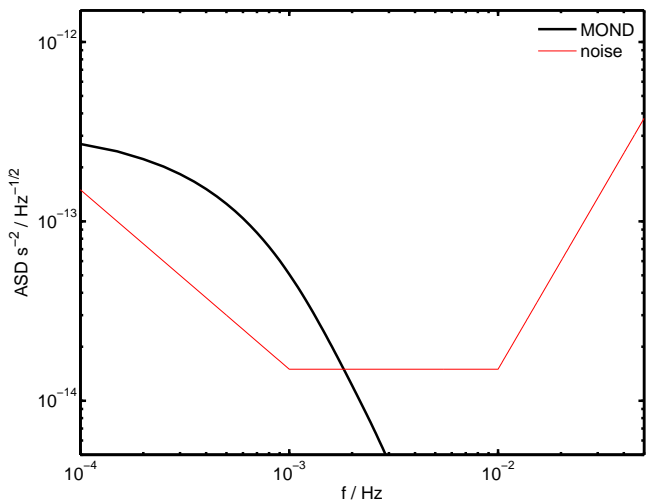}}\\
\caption{\label{fig:stressNoise1}The amplitude spectral density (ASD) for the $S_{zz}$ MOND signal for a $y=50\;\unit{km}$, $\dot{x}=1\;\unit{kms^{-1}}$ trajectory, compared to the simple noise model employed here. Note that while the noise ASD is independent of the integration time, there being a single MOND event means that its ASD lessens with increasing integration time, where for the present plot we conservatively integrate over $2\times10^4\unit{s}$.}
\end{figure}

The results in \Fig{stressNoise} assume a spacecraft velocity of $1\unit{kms^{-1}}$ through the Earth-Sun saddle region, passing on a trajectory with $y=50\;\unit{km}$ and $z=0$, during a full Moon. While the velocity of the probe is unlikely to be aligned with the $x$-axis in this manner, the details of the trajectory orientation do not appear to be greatly important. A significant signal is clearly seen in both real-time and Fourier domains: Figs. \ref{fig:stressNoise} and \ref{fig:stressNoise1}, respectively. If the actual trajectory saw LPF cross the saddle region more quickly, then the MOND signal would be shifted to higher frequencies where the signal to noise is even better. An increase in the impact parameter $r_{\min}$ reduces the signal roughly as $1/r_{\min}$ for $r_{\min}\sim r_0$, but as $1/r_{\min}^{2}$ for $r_{\min}\gg r_0$. 



\section{Conclusions}
\label{concs}

We solved the non-relativistic limit of T$e$V$e$S for two 
and three-body saddle points
using a particular $\mu$ function used previously in the literature. We predicted an anomalous stress signal that is potentially detectable by future space-probes. We found that past analytical calculations for the two-body case are not only in agreement with our numerical results, at least in their regimes of applicability, but provide a good insight into three-body problems of the Sun-Earth-Moon type, where the Newtonian stresses at the saddle points are dominated by a single body. Our results indicate that the most significant effect of the Moon on the Earth-Sun saddle is to change its location, while the MOND anomalous stress signal is not greatly affected, and is stable to changes in the phase of Moon. Furthermore, we have demonstrated that the single lunar saddle point near the Moon yields a stress signal of similar form but differs in stress magnitude and spatial extent. However, for an impact parameter likely from a spacecraft flyby, we note that these two factors effectively cancel (at least in some regime).

Finally, we have demonstrated that a possible extension to the LISA Pathfinder mission, targeting the saddles, would be sensitive to the anomalous stress signal that we have calculated. This is in part due to the distinct form of the signal close to a saddle point, which cannot be confused with the (much smoother) Newtonian signal (or with any other signal derived from the parameterized post-Newtonian framework). It is also due to the likely speed of the probe through the saddle region yielding signal variations at frequencies that minimize the instrument noise. While this conclusion assumes the ability to pass within $\sim100\unit{km}$ of a saddle point, we believe that this is achievable even with the low thrust propulsion system incorporated in the LISA Pathfinder design.  In future work we hope to implement our studies as part of a concerted effort to explore realistic trajectories for LISA Pathfinder.

We conclude with a general remark on MONDian theories which we hope
will clarify a number of issues with the proposed test. 
For all MONDian theories with a relativistic formulation
(and not only TeVeS, as described in the introduction),
MONDian non-relativistic effects are due to an extra field, $\phi$, as 
opposed to the gravitational field itself, i.e. the metric field, with $\Phi_N$
associated with $g_{00}$. Indeed it's
very hard to covariantly modify a gravity theory so that in the 
non-relativistic limit $g_{00}$ is ruled by a non-linear Poisson equation.
In the Newtonian limit (when $\mu\approx 1$ in Eqn.(\ref{eqn:TeVeS})) 
the field $\phi$ mimics the Newtonian 
field and so renormalizes the gravitational constant ($G\rightarrow
G(1+\kappa/(4\pi)$) as discussed
in the literature (e.g.~\cite{teves}). To satisfy constraints, 
$\phi$ must be sub-dominant in this regime, and this is enforced
by the parameter  $\kappa$ appearing in its Poisson equation 
(\ref{eqn:TeVeS}).

But this simple fact implies that one must trigger MONDian behavior in $\phi$ 
at Newtonian accelerations $a_N$ much higher than $a_0$.
Only thus can $\phi$'s relative importance start to increase with decreasing 
$a_N$, so that by the time $a_N\sim a_0$ the field $\phi$ is not only 
MONDian but dominates $\Phi_N$, as required by astronomical
applications. It can be easily computed that 
if $\mu$ turns from 1 to a single power-law, then MONDian behavior in 
$\phi$ should be triggered when the Newtonian acceleration $a_N$ drops below  
$(4\pi/\kappa)^2 a_0\sim 1.75 \times 10^5 a_0\sim 10^{-5} \unit{m s}^{-2}$. 
And indeed this  is the rough Newtonian acceleration at $r_0$. 
It is therefore important
to realize that for LPF realistic impact parameters we would probe the regime
where $\phi$ has gone fully MONDian, but hasn't yet dominated $\Phi_N$.
The MOND signal can be detected above the Newtonian one because it has a 
distinctive spatial variation whereas $\Phi_N$ is just a DC component.

Bearing this in mind a number of issues may be clarified. One concerns
the self-gravity of LPF. This is only balanced at the level of $a_N\sim 10^{-9}
\unit{ms}^{-2}$, so one might think that a test of MOND, in the regime $a_N
\sim a_0$, would run against the wall of 
self-gravity. In fact we are testing much higher
Newtonian accelerations; for instance, for an impact parameter of 40~Km
we have $a_N\sim 10^{-6}\unit{ms}^{-2}$. 
We'd need to approach the saddle much closer than
about $400$ meters before self-gravity became an issue (and the spacecraft
itself had to be included in the computation of the location of the saddle).

This should also  clear the matter of the generality of the predictions
made. There are several $\mu$ on offer for astrophysical
purposes. They all have the same rough behavior in the regime under study,
where $\phi$ is fully MONDian but hasn't yet dominated $\Phi_N$. 
A negative result from LPF would rule out virtually
all proposed $\mu$. It would only fail to rule out very contrived
$\mu(y)$ (never suggested in the literature), for which two power laws are 
employed in $\mu(y)$: a very steep one 
from $y=1$ (felt at $r_0$) to the point where $a_N=a_0$;
and the usual linear power-law for $a_N<a_0$:
extremely contrived. We are currently writing a follow up paper expanding
on this matter,

The high sensitivity of LISA Pathfinder enables it to potentially detect the anomalous MOND stress when the deviations from Newtonian dynamics are tiny, in contrast to the conditions present in outer reaches of galaxies~\cite{binney,yusaf}. On the other hand, constraints are yielded by planetary orbits~\cite{Sereno} only at much stronger accelerations than LISA Pathfinder would probe. A saddle point fly-by offers the chance to study MOND where we have an exquisite knowledge of the mass distribution and the ability to make full calculations of the MOND acceleration field, without assuming simplifying symmetries. Furthermore, saddle point exploration would enable an investigation of gravity at very low accelerations in the close vicinity of Earth, which can be reached in a short time-frame and without the increased mass distribution uncertainties affecting probes sent to very large distances from the Sun. A positive detection would, of course, vindicate the MOND paradigm, while a null result would provide clean and reliable constraints upon it.

In closing we note that several refinements to our numerical calculations
should be trivial. Including the quadrupole and higher multipole moments
of the sources or the effect of Jupiter and other Solar system objects
(or even the galactic field)
is straightforward, as they only affect the boundary conditions for our problem.
However these effects are expected to be small. The shift of the saddle
location is of course non-negligible, but can be easily computed 
with effects already studied 
in~\cite{Bekenstein:2006fi}. Once the new location is taken into account
the change in the detailed MOND tidal stresses can be
computed just by changing the boundary conditions of our code.
However, if the effect of the Moon on the Earth-Sun predictions for realistic
impact parameters is already so small, these effects can be expected to be 
completely negligible.

\begin{acknowledgments}
We acknowledge financial support from STFC (N.B.) and thank J. Bekenstein, 
and  
C. Skordis   for helpful discussions. The numerical
work was performed on the COSMOS supercomputer (which is
supported by STFC, HEFCE and SGI) and the Imperial College HPC facilities.
\end{acknowledgments}


\appendix

\section{Details of numerical method}\label{sec:numerics}
Here we briefly describe our numerical algorithm. This is based upon representing $\vect{g}$ on the sites of a non-uniform lattice of the form shown in \Fig{grid}, such that we obtain greater spatial resolution near the saddle point. A relaxation algorithm then cycles over each lattice site $\x$ and changes the values of $\vect{g}$ at $\x$ and the neighboring sites so that the divergence equation \Eq{laplace} is solved \emph{locally}, that is:
\begin{equation}
\label{eqn:solveLocal}
D_{\x} = \sum_{j}\left( \mu_{\x} g_{\x}^{j} - \mu_{\xminus{j}} g_{\xminus{j}}^{j} \right) = 0.
\end{equation}
We then move to the next site and change the field values so that the condition is valid there, using the newly updated values as we proceed. However, when enforcing the above condition at these subsequent sites, the value of $D_{\x}$ at the first site will be slightly changed. We therefore require many cycles over the whole lattice before the above condition is closely matched globally. 

We stress that we use a different discretization procedure to Ref.~\cite{Milgrom:1986ib}, in that we define all three components of $\vect{g}$ as well as $\mu$ at the same location, that is:
\begin{equation}
\mu_{\x} = \mu( \kappa \vect{g}_{\x} / a_{0}).
\end{equation}
This is in contrast to the more elaborate scheme used in Ref.~\cite{Milgrom:1986ib}, in which $u^{j}$ is defined on the $j$-links between the lattice sites and then $\mu$ is defined at the centers of the grid squares in the 2D case considered. Under that scheme the calculation of $u^{i}/\mu$ at any position in a 3D calculation would require knowledge of $33$ values of $u^{j}$, whereas to determine $\mu_{\x} g^{j}_{\x}$ here, we require merely the knowledge of the three components of $\vect{g}_{\x}$.

The presence of the non-linear function $\mu$ is a notable complication. Furthermore its form is not uniquely known and therefore we desire to be able to solve for an arbitrary function. The algorithm of Ref.~\cite{Milgrom:1986ib} involves solving the curl equation for $\vect{u}$ locally, while keeping the $\mu$ values fixed at their old values. Once the $\mu$ values are updated using the new $\vect{u}$, the curl equation is no longer matched, and this slows the convergence of the algorithm. In contrast, here we proceed by solving the divergence equation to first order in $\delta\vect{g}$ and $\delta\mu$, where $\delta$ denotes the change from one step to the next. Then, as the system converges to the solution, the terms of order $\delta^{2}$ become negligible very rapidly. 

Crucial in the above is to ensure that in updating the field configuration
the curl of $\vect{g}$ remains zero. We define the our preserved discrete curl as:
\begin{equation}
(\vect{\Del}\times\vect{g})^{k}_{\x}
= 
\frac{g^{j}_{\xplus{i}}-g^{j}_{\x}}{r^{i}_{\xplus{i}}-r^{i}_{\x}}
-
\frac{g^{i}_{\xplus{j}}-g^{i}_{\x}}{r^{j}_{\xplus{j}}-r^{j}_{\x}},
\end{equation}
where $\vect{r}_{\x}$ is the position vector site $\x$. In order to preserve this discrete curl at each step of the relaxation we change the fields according to:
\begin{eqnarray}
\label{eqn:preserveCurl1}
\delta g_{\x}^{j} = \frac{+C_{\x}}{r^{j}_{\xplus{j}}-r^{j}_{\x}}, \\
\label{eqn:preserveCurl2}
\delta g_{\xminus{j}}^{j} = \frac{-C_{\x}}{r^{j}_{\x}-r^{j}_{\xminus{j}}}, 
\end{eqnarray}
where the value of $C_{\x}$ is chosen, as below, to yield \Eq{solveLocal} to first order. While the first equation acts on all three components of $\vect{g}$, the second acts on only one component per site. 

To determined $C_{\x}$, consider that to first order in $\delta\vect{g}$ and $\delta\mu$:
\begin{eqnarray}
\delta D_{\x}
\approx 
\sum_{j} 
\frac
 {\mu_{\x} \delta g^{j}_{\x} + g^{j}_{\x} \delta \mu_{\x}
   - \mu_{\xminus{j}} \delta g^{j}_{\xminus{j}} - g^{j}_{\xminus{j}} \delta \mu_{\xminus{j}}}
 {r^{j}_{\x}-r^{j}_{\xminus{j}}}.
\end{eqnarray}
Writing $\delta\mu$ in terms of $\delta g^{2}$ and substituting the above values for $\delta\vect{g}$ then yields:
\begin{eqnarray}
\frac{\delta D_{\x}}{C_{\x}} 
\approx
\sum_{j}\left[
 \frac{\mu_{\x}}{\Delta^{j}_{-}\Delta^{j}_{+}}
 + 2 \sum_{i} \frac{g^{i}_{\x}}{\Delta^{i}_{+}} 
              \frac{g^{j}_{\x}}{\Delta^{j}_{-}}
              \frac{\diff\mu_{\x}}{\diff g^{2}_\vect{\x}} \right.\\
 \left. + \;
     \frac{\mu_{\xminus{j}}}{ (\Delta^{j}_{-} )^{2}}
 \; + \; 
     2 \left( \frac{g^{j}_{\xminus{j}}}{\Delta^{j}_{-}} \right)^{\!\!\!2}
     \frac{\diff\mu_{\xminus{j}}}{\diff g^{2}_\vect{\xminus{j}}}\right]
\end{eqnarray}
where we have used compact notation such that:
\begin{eqnarray}
\Delta^{j}_{+} & = & r^{j}_{\xplus{j}} - r^{j}_{\x}
\\
\Delta^{j}_{-} & = & r^{j}_{\x} - r^{j}_{\xminus{j}}.
\end{eqnarray}
We can then insert the $\mu$ derivatives for our chosen $\mu$ function. Finally, since we want $D_{\x}$ to be zero after each change, we set:
\begin{equation}
\delta D_{\x} = - D_{\x},
\end{equation}
and hence know the $C_{\x}$ required to obtain $D_{\x}=0$, at least to first order. (If $\mu$ had not changed during this step, the above procedure would have set $D_{\x}$ to be exactly zero, i.e. to all orders in $\delta\vect{g}$.)

Note that in addition to preserving the curl, Eqs. (\ref{eqn:preserveCurl1}) and (\ref{eqn:preserveCurl2}) also exactly preserve the change in $\PHI$ measured across the lattice as:
\begin{equation}
\Delta \PHI = \sum_{x^{j}} \left( r^{j}_{\xplus{j}} - r^{j}_{\x} \right) g^{j}_{\x} ,
\end{equation}
where $x^{i}$ and $x^{k}$ are fixed during the summation. 

In practice, cycling over the lattice and solving the discrete equation locally does not lead to rapid enough convergence to the solution. As $D_{x}$ is (approximately) zeroed at later sites, $D_{x}$ at earlier sites is moved slightly away from its desired value and a large number of iterations of this procedure are required before each $\vect{g}_{\x}$ has converged to a good approximation. We therefore preempt the changes to the field that will occur at other points in the cycle using the fact that these very changes are (largely) responsible for $D_{\x}$ being non-zero. This is achieved by a method known as successive over-relaxation (SOR, e.g.~\cite{NumRecC}) in which $\delta g^{j} \rightarrow \lambda\delta g^{j}$, where $\lambda$ is the over-relaxation parameter and is larger than unity. We begin with $\lambda=1$ and increase it once the system has begun to settle down, since high values of $\lambda$ can initially result in the RMS value of $D_{x}$ increasing, contrary to our goal.

We coded the algorithm outlined in this Appendix using the LATfield library~\cite{latfield}.

\bibliography{references}

\begin{thebibliography}{28}
\expandafter\ifx\csname natexlab\endcsname\relax\def\natexlab#1{#1}\fi
\expandafter\ifx\csname bibnamefont\endcsname\relax
  \def\bibnamefont#1{#1}\fi
\expandafter\ifx\csname bibfnamefont\endcsname\relax
  \def\bibfnamefont#1{#1}\fi
\expandafter\ifx\csname citenamefont\endcsname\relax
  \def\citenamefont#1{#1}\fi
\expandafter\ifx\csname url\endcsname\relax
  \def\url#1{\texttt{#1}}\fi
\expandafter\ifx\csname urlprefix\endcsname\relax\def\urlprefix{URL }\fi
\providecommand{\bibinfo}[2]{#2}
\providecommand{\eprint}[2][]{\url{#2}}

\bibitem[{\citenamefont{Milgrom}(1983)}]{Milgrom:1983ca}
\bibinfo{author}{\bibfnamefont{M.}~\bibnamefont{Milgrom}},
  \bibinfo{journal}{Astrophys. J.} \textbf{\bibinfo{volume}{270}},
  \bibinfo{pages}{365} (\bibinfo{year}{1983}).

\bibitem[{\citenamefont{Bekenstein and Milgrom}(1984)}]{aqual}
\bibinfo{author}{\bibfnamefont{J.}~\bibnamefont{Bekenstein}} \bibnamefont{and}
  \bibinfo{author}{\bibfnamefont{M.}~\bibnamefont{Milgrom}},
  \bibinfo{journal}{Astrophys. J.} \textbf{\bibinfo{volume}{286}},
  \bibinfo{pages}{7} (\bibinfo{year}{1984}).

\bibitem[{\citenamefont{Bekenstein}(2005)}]{teves}
\bibinfo{author}{\bibfnamefont{J.~D.} \bibnamefont{Bekenstein}},
  \bibinfo{journal}{{Phys. Rev. \textbf{D70}, 083509 (2004); Erratum-ibid.}}
  \textbf{\bibinfo{volume}{D71}}, \bibinfo{pages}{069901}
  (\bibinfo{year}{2005}), \eprint{astro-ph/0403694}.

\bibitem[{\citenamefont{Sanders}(2005)}]{BSTV}
\bibinfo{author}{\bibfnamefont{R.~H.} \bibnamefont{Sanders}},
  \bibinfo{journal}{Mon. Not. Roy. Astron. Soc.}
  \textbf{\bibinfo{volume}{363}}, \bibinfo{pages}{459} (\bibinfo{year}{2005}),
  \eprint{astro-ph/0502222}.

\bibitem[{\citenamefont{Zlosnik et~al.}(2006)\citenamefont{Zlosnik, Ferreira,
  and Starkman}}]{aether}
\bibinfo{author}{\bibfnamefont{T.~G.} \bibnamefont{Zlosnik}},
  \bibinfo{author}{\bibfnamefont{P.~G.} \bibnamefont{Ferreira}},
  \bibnamefont{and} \bibinfo{author}{\bibfnamefont{G.~D.}
  \bibnamefont{Starkman}}, \bibinfo{journal}{Phys. Rev.}
  \textbf{\bibinfo{volume}{D74}}, \bibinfo{pages}{044037}
  (\bibinfo{year}{2006}), \eprint{gr-qc/0606039}.

\bibitem[{\citenamefont{Zlosnik et~al.}(2007)\citenamefont{Zlosnik, Ferreira,
  and Starkman}}]{aether1}
\bibinfo{author}{\bibfnamefont{T.~G.} \bibnamefont{Zlosnik}},
  \bibinfo{author}{\bibfnamefont{P.~G.} \bibnamefont{Ferreira}},
  \bibnamefont{and} \bibinfo{author}{\bibfnamefont{G.~D.}
  \bibnamefont{Starkman}}, \bibinfo{journal}{Phys. Rev.}
  \textbf{\bibinfo{volume}{D75}}, \bibinfo{pages}{044017}
  (\bibinfo{year}{2007}), \eprint{astro-ph/0607411}.

\bibitem[{\citenamefont{Ferreras et~al.}(2008)\citenamefont{Ferreras,
  Sakellariadou, and Yusaf}}]{yusaf}
\bibinfo{author}{\bibfnamefont{I.}~\bibnamefont{Ferreras}},
  \bibinfo{author}{\bibfnamefont{M.}~\bibnamefont{Sakellariadou}},
  \bibnamefont{and} \bibinfo{author}{\bibfnamefont{M.~F.} \bibnamefont{Yusaf}},
  \bibinfo{journal}{Phys. Rev. Lett.} \textbf{\bibinfo{volume}{100}},
  \bibinfo{pages}{031302} (\bibinfo{year}{2008}), \eprint{0709.3189}.

\bibitem[{\citenamefont{Mavromatos et~al.}(2009)\citenamefont{Mavromatos,
  Sakellariadou, and Yusaf}}]{yusaf1}
\bibinfo{author}{\bibfnamefont{N.~E.} \bibnamefont{Mavromatos}},
  \bibinfo{author}{\bibfnamefont{M.}~\bibnamefont{Sakellariadou}},
  \bibnamefont{and} \bibinfo{author}{\bibfnamefont{M.~F.} \bibnamefont{Yusaf}},
  \bibinfo{journal}{Phys. Rev.} \textbf{\bibinfo{volume}{D79}},
  \bibinfo{pages}{081301} (\bibinfo{year}{2009}), \eprint{0901.3932}.

\bibitem[{\citenamefont{Ferreras et~al.}(2009)\citenamefont{Ferreras,
  Mavromatos, Sakellariadou, and Yusaf}}]{yusaf2}
\bibinfo{author}{\bibfnamefont{I.}~\bibnamefont{Ferreras}},
  \bibinfo{author}{\bibfnamefont{N.~E.} \bibnamefont{Mavromatos}},
  \bibinfo{author}{\bibfnamefont{M.}~\bibnamefont{Sakellariadou}},
  \bibnamefont{and} \bibinfo{author}{\bibfnamefont{M.~F.} \bibnamefont{Yusaf}},
  \bibinfo{journal}{Phys. Rev.} \textbf{\bibinfo{volume}{D80}},
  \bibinfo{pages}{103506} (\bibinfo{year}{2009}), \eprint{0907.1463}.

\bibitem[{\citenamefont{Skordis}(2009)}]{kostasrev}
\bibinfo{author}{\bibfnamefont{C.}~\bibnamefont{Skordis}},
  \bibinfo{journal}{Class. Quant. Grav.} \textbf{\bibinfo{volume}{26}},
  \bibinfo{pages}{143001} (\bibinfo{year}{2009}), \eprint{0903.3602}.

\bibitem[{\citenamefont{Clowe et~al.}(2006)}]{bullet}
\bibinfo{author}{\bibfnamefont{D.}~\bibnamefont{Clowe}} \bibnamefont{et~al.},
  \bibinfo{journal}{Astrophys. J.} \textbf{\bibinfo{volume}{648}},
  \bibinfo{pages}{L109} (\bibinfo{year}{2006}), \eprint{astro-ph/0608407}.

\bibitem[{\citenamefont{Dai et~al.}(2008)\citenamefont{Dai, Matsuo, and
  Starkman}}]{bullet1}
\bibinfo{author}{\bibfnamefont{D.-C.} \bibnamefont{Dai}},
  \bibinfo{author}{\bibfnamefont{R.}~\bibnamefont{Matsuo}}, \bibnamefont{and}
  \bibinfo{author}{\bibfnamefont{G.}~\bibnamefont{Starkman}},
  \bibinfo{journal}{Phys. Rev.} \textbf{\bibinfo{volume}{D78}},
  \bibinfo{pages}{104004} (\bibinfo{year}{2008}), \eprint{0806.4319}.

\bibitem[{\citenamefont{Angus and McGaugh}(2007)}]{bullet2}
\bibinfo{author}{\bibfnamefont{G.~W.} \bibnamefont{Angus}} \bibnamefont{and}
  \bibinfo{author}{\bibfnamefont{S.~S.} \bibnamefont{McGaugh}}
  (\bibinfo{year}{2007}), \eprint{0704.0381}.

\bibitem[{\citenamefont{Brownstein and Moffat}(2007)}]{bullet3}
\bibinfo{author}{\bibfnamefont{J.~R.} \bibnamefont{Brownstein}}
  \bibnamefont{and} \bibinfo{author}{\bibfnamefont{J.~W.}
  \bibnamefont{Moffat}}, \bibinfo{journal}{Mon. Not. Roy. Astron. Soc.}
  \textbf{\bibinfo{volume}{382}}, \bibinfo{pages}{29} (\bibinfo{year}{2007}),
  \eprint{astro-ph/0702146}.

\bibitem[{\citenamefont{Sereno and Jetzer}(2006)}]{Sereno}
\bibinfo{author}{\bibfnamefont{M.}~\bibnamefont{Sereno}} \bibnamefont{and}
  \bibinfo{author}{\bibfnamefont{P.}~\bibnamefont{Jetzer}},
  \bibinfo{journal}{Mon. Not. Roy. Astron. Soc.}
  \textbf{\bibinfo{volume}{371}}, \bibinfo{pages}{626} (\bibinfo{year}{2006}),
  \eprint{astro-ph/0606197}.

\bibitem[{\citenamefont{Turyshev and Toth}(2009)}]{pioneer0}
\bibinfo{author}{\bibfnamefont{S.~G.} \bibnamefont{Turyshev}} \bibnamefont{and}
  \bibinfo{author}{\bibfnamefont{V.~T.} \bibnamefont{Toth}}
  (\bibinfo{year}{2009}), \eprint{0906.0399}.

\bibitem[{\citenamefont{Brownstein and Moffat}(2006)}]{pioneer}
\bibinfo{author}{\bibfnamefont{J.~R.} \bibnamefont{Brownstein}}
  \bibnamefont{and} \bibinfo{author}{\bibfnamefont{J.~W.}
  \bibnamefont{Moffat}}, \bibinfo{journal}{Class. Quant. Grav.}
  \textbf{\bibinfo{volume}{23}}, \bibinfo{pages}{3427} (\bibinfo{year}{2006}),
  \eprint{gr-qc/0511026}.

\bibitem[{\citenamefont{Bekenstein and Magueijo}(2006)}]{Bekenstein:2006fi}
\bibinfo{author}{\bibfnamefont{J.}~\bibnamefont{Bekenstein}} \bibnamefont{and}
  \bibinfo{author}{\bibfnamefont{J.}~\bibnamefont{Magueijo}},
  \bibinfo{journal}{Phys. Rev.} \textbf{\bibinfo{volume}{D73}},
  \bibinfo{pages}{103513} (\bibinfo{year}{2006}), \eprint{astro-ph/0602266}.

\bibitem[{\citenamefont{Milgrom}(2009)}]{Milgromss}
\bibinfo{author}{\bibfnamefont{M.}~\bibnamefont{Milgrom}}
  (\bibinfo{year}{2009}), \eprint{0906.4817}.

\bibitem[{\citenamefont{McNamara et~al.}(2008)\citenamefont{McNamara, Vitale,
  and Danzmann}}]{LISA}
\bibinfo{author}{\bibfnamefont{P.}~\bibnamefont{McNamara}},
  \bibinfo{author}{\bibfnamefont{S.}~\bibnamefont{Vitale}}, \bibnamefont{and}
  \bibinfo{author}{\bibfnamefont{K.}~\bibnamefont{Danzmann}}
  (\bibinfo{collaboration}{LISA}), \bibinfo{journal}{Class. Quant. Grav.}
  \textbf{\bibinfo{volume}{25}}, \bibinfo{pages}{114034}
  (\bibinfo{year}{2008}).

\bibitem[{\citenamefont{Milgrom}(1986)}]{Milgrom:1986ib}
\bibinfo{author}{\bibfnamefont{M.}~\bibnamefont{Milgrom}},
  \bibinfo{journal}{Astrophys. J.} \textbf{\bibinfo{volume}{302}},
  \bibinfo{pages}{617} (\bibinfo{year}{1986}).

\bibitem[{\citenamefont{Famaey and Binney}(2005)}]{binney}
\bibinfo{author}{\bibfnamefont{B.}~\bibnamefont{Famaey}} \bibnamefont{and}
  \bibinfo{author}{\bibfnamefont{J.}~\bibnamefont{Binney}},
  \bibinfo{journal}{Mon. Not. Roy. Astron. Soc.}
  \textbf{\bibinfo{volume}{363}}, \bibinfo{pages}{603} (\bibinfo{year}{2005}),
  \eprint{astro-ph/0506723}.

\bibitem[{\citenamefont{Nipoti et~al.}(2008)\citenamefont{Nipoti, Londrillo,
  and Ciotti}}]{Cioti}
\bibinfo{author}{\bibfnamefont{C.}~\bibnamefont{Nipoti}},
  \bibinfo{author}{\bibfnamefont{P.}~\bibnamefont{Londrillo}},
  \bibnamefont{and} \bibinfo{author}{\bibfnamefont{L.}~\bibnamefont{Ciotti}}
  (\bibinfo{year}{2008}), \eprint{0811.2878}.

\bibitem[{\citenamefont{Tiret and Combes}(2007)}]{Tiret}
\bibinfo{author}{\bibfnamefont{O.}~\bibnamefont{Tiret}} \bibnamefont{and}
  \bibinfo{author}{\bibfnamefont{F.}~\bibnamefont{Combes}}
  (\bibinfo{year}{2007}), \eprint{astro-ph/0701011}.

\bibitem[{\citenamefont{Bevis et~al.}(2010)\citenamefont{Bevis, Magueijo, and
  Mozzafari}}]{future}
\bibinfo{author}{\bibfnamefont{N.}~\bibnamefont{Bevis}},
  \bibinfo{author}{\bibfnamefont{J.}~\bibnamefont{Magueijo}}, \bibnamefont{and}
  \bibinfo{author}{\bibfnamefont{A.}~\bibnamefont{Mozzafari}},
  \bibinfo{journal}{in preparation}  (\bibinfo{year}{2010}).

\bibitem[{\citenamefont{Trenkel et~al.}(2009)\citenamefont{Trenkel, Kemble,
  Bevis, and Magueijo}}]{companion}
\bibinfo{author}{\bibfnamefont{C.}~\bibnamefont{Trenkel}},
  \bibinfo{author}{\bibfnamefont{S.}~\bibnamefont{Kemble}},
  \bibinfo{author}{\bibfnamefont{N.}~\bibnamefont{Bevis}}, \bibnamefont{and}
  \bibinfo{author}{\bibfnamefont{J.}~\bibnamefont{Magueijo}},
  \bibinfo{journal}{submitted}  (\bibinfo{year}{2009}).

\bibitem[{\citenamefont{{Press, William and al.}}(1992)}]{NumRecC}
\bibinfo{author}{\bibnamefont{{Press, William and al.}}}
  (\bibinfo{year}{1992}), \bibinfo{note}{{Numerical Recipes in C, Cambridge
  University Press, UK.}}

\bibitem[{\citenamefont{Bevis and Hindmarsh}()}]{latfield}
\bibinfo{author}{\bibfnamefont{N.}~\bibnamefont{Bevis}} \bibnamefont{and}
  \bibinfo{author}{\bibfnamefont{M.}~\bibnamefont{Hindmarsh}},
  \urlprefix\url{www.latfield.org}.

\end{thebibliography}

\end{document}